\documentclass[universe,article,accept,pdftex,moreauthors]{Definitions/mdpi} 

\usepackage{natbib}
\usepackage{amssymb,amsmath}
\usepackage{bm}
\usepackage{siunitx}
\usepackage{longtable}
\usepackage{subcaption}
\usepackage{cancel}
\usepackage{xcolor}
\usepackage[normalem]{ulem}

\usepackage[final]{changes}

\firstpage{1} 
\pubvolume{11}
\issuenum{6}
\articlenumber{178}
\pubyear{2025}
\copyrightyear{2025}
\externaleditor{Panayiotis \\Stavrinos}
\datereceived{19 April 2025} 
\daterevised{27 May 2025} 
\dateaccepted{29 May 2025} 
\datepublished{31 May 2025} 

\hreflink{https://doi.org/10.3390/\linebreak universe11060178} 

\Title{Statistical Strong Lensing as a Test of Conformal Gravity}

\TitleCitation{Statistical Strong Lensing as a Test of Conformal Gravity}

\Author{ {Li-Xue Yue} 
	$^{1,2,}$* and Da-Ming Chen $^{1,2}$\orcidA{}}

\AuthorNames{Li-Xue Yue and Da-Ming Chen}

\AuthorCitation{Yue, L.-X.; Chen, D.-M.}

\address{%
	$^{1}$ \quad National Astronomical Observatories, Chinese Academy of Sciences, Beijing 100101, China; cdm@nao.cas.cn\\
	$^{2}$ \quad School of Astronomy and Space Science, University of Chinese Academy of Sciences, Beijing 100049, China}

\corres{Correspondence: lxyue@nao.cas.cn}

\abstract{As an alternative gravitational theory to General Relativity (GR), Conformal Gravity (CG) can be verified through astronomical observations. Currently, {Mannheim and Kazanas have provided vacuum solutions for cosmological and local gravitational systems, and these solutions may resolve the dark matter and dark energy issues encountered in GR, making them particularly valuable}. {For static, spherically symmetric systems,} CG predicts an additional linear potential generated by luminous matter in addition to the conventional Newtonian potential. This extra potential is expected to account for the observations of galaxies and galaxy clusters without the need of dark matter. It is characterized by the parameter $\gamma^*$, which {corresponds to the linear potential generated by the unit of the solar mass, and it is thus a universal constant}. The value of $\gamma^*$ was determined by fitting the rotation curve data of spiral galaxies. These predictions of CG should also be verified by the observations of strong gravitational lensing. To date, in the existing literature, the observations of strong lensing employed to test CG have been limited to a few galaxy clusters. It has been found that the value of $\gamma^*$ estimated from strong lensing is several orders of magnitude greater than that obtained from fitting rotation curves. In this study, building upon the previous research, we tested CG via strong lensing {statistics}. We used a well-defined sample that consisted of both galaxies and galaxy clusters. This allowed us to test CG through statistical strong lensing in a way similar to the conventional approach in GR. As anticipated, our results were consistent with previous studies, {namely that the fitted $\gamma^*$ is much larger than that from rotation curves}. Intriguingly, we further discovered that, in order to fit the strong lensing data {of another sample}, the value of $\gamma^*$ cannot be a constant, as is required in CG. Instead, we derived a formula for $\gamma^*$ as a function of the stellar mass $M_*$ of the galaxies or galaxy clusters. It was found that $\gamma^*$ decreases as \mbox{$M_*$ increases}.}


\AtBeginDocument{
	\DeclareUnicodeCharacter{22C6}{$^\star$} 
}

\graphicspath{{./}}

\begin{document}
	
	\section{Introduction} \label{sec:introduction}
	According to General Relativity (GR), gravity can be characterized by the curvature of spacetime. The~curvature of spacetime is dictated by the matter distribution within it. Test objects, and~even light, traverse along the geodesics in the curved spacetime. As~a result, if~light passes near a massive object, such as a star, a~galaxy, or a galaxy cluster, the~bending of light will occur. This phenomenon is known as gravitational lensing.
	Now, gravitational lensing has evolved into an independent and powerful tool for constraining or examining the matter distribution profile and cosmological parameters in the standard $\Lambda$CDM cosmology, as~cited in \citep{schneider1999gravitational}. On~the other hand, GR turns out to be accurate only at the scales of a solar system. This includes its successful prediction of the deflection angle when distant light rays pass the edge of the sun. This prediction, also known as one of the three classic tests of GR, represents the earliest successful application of gravitational lensing~\citep{Weinberg:1972kfs}. At~larger scales, including the whole universe,  dark matter (DM) and dark energy (DE) must be introduced to ensure that GR remains a valid theory of gravity. However, as~DM and DE still lack credible theoretical underpinnings and direct empirical evidence, numerous alternative gravity theories to GR have been developed. These alternative theories must undergo the scrutiny of astronomical and cosmological observations, including gravitational lensing. Conformal Gravity (CG)~\citep{MANNHEIM2006340} is one such alternative~theory.
	
	Conformal Gravity (CG) has gained increasing attention as a potential alternative to dark matter (DM) and dark energy (DE), with~its predictions being actively tested against astronomical observations \citep{MANNHEIM2006340}. As~a relativistic theory extending beyond General Relativity (GR), CG offers solutions to both the cosmological constant problem inherent in $\Lambda$CDM cosmology \citep{1992ApJ...391..429M,Mannheim2000FoPh,Mannheim_2001} and it demonstrates consistency with Type Ia supernova data \citep{MANNHEIM2006340,YANG201343}.
	
	In the non-relativistic limit, CG predicts a modified linear gravitational potential that supplements the Newtonian potential \citep{1989ApJ...342..635M}, potentially explaining galactic rotation curves without invoking DM. Recent studies have extensively tested this framework through detailed fits to observed the rotation curves of spiral galaxies \citep{mannheim_fitting_2012,Mannheim_2013,2015JPhCS.615a2002O}. These results consistently indicate a universal constant $\gamma^\ast=5.42 \times 10^{-39}$ m$^{-1}$, corresponding to the linear potential generated by a solar--mass ($M_\odot$) point source. For~an arbitrary mass $M$, the~linear potential scales as $\gamma=(M/M_\odot)\gamma^*$, where $\gamma$ parametrizes the strength of this~contribution.
	
	Strong gravitational lensing has emerged as a valuable test for Conformal Gravity (CG), attracting growing research interest. While several studies \citep{PhysRevD.58.024011,PhysRevD.81.127502,sultana2013deflection,PhysRevD.87.047503,PhysRevD.95.024004,PhysRevD.100.024019} have derived light deflection angle expressions and performed lensing analyses in Mannheim--Kazanas spacetimes, the~reported bending angle formulae show significant discrepancies and fail to converge. To~date, tests of Conformal Gravity (CG) through strong lensing have been limited to individual galaxy clusters. In~this work, we expanded the analysis to some well-defined samples of both galaxies and galaxy clusters. This approach enables systematic constraints on the $\gamma^*$ parameter via strong lensing statistics, while probing for potential correlations between $\gamma^*$ and the lens mass ($M$).
	
	This paper is organized as follows: Section~\ref{sec:CG introduction} reviews key Conformal Gravity (CG) results relevant to our analysis. Section~\ref{sec:deflection angle} derives the CG deflection angle for lensing systems. In~Section~\ref{sec:Fitting the gamma^*}, the~parameter \( \gamma^* \) is re-fitted using strong gravitational lensing data.
	We then present the galaxy stellar mass function in Section~\ref{sec:GSMF}, which is followed by the lensing probability distribution fits for $\gamma^*$ in Section~\ref{sec:lensing probability}. Finally, Section~\ref{sec:conclusions} provides the conclusions and~discussion.
	
	\section{Cosmology and Galacdynamics in~CG} \label{sec:CG introduction}
	Similar to General Relativity (GR), Conformal Gravity (CG) is formulated by taking the metric $g_{\mu\nu}$ as the gravitational field. Nevertheless, it bestows an extra local symmetry upon gravity, namely the conformal symmetry $g_{\mu \nu}(x) \to e^{2\alpha(x)} g_{\mu \nu}(x)$, which surpasses the ordinary coordinate invariance. The~Weyl tensor  $C^{\lambda\mu\nu\kappa}$, defined by~\citep{MANNHEIM2006340},
	\begin{equation}\label{eq:Weyl tensor}
		C_{\lambda\mu\nu\kappa} = R_{\lambda\mu\nu\kappa} - \frac{1}{2} \left(g_{\lambda\nu} R_{\mu\kappa} - g_{\lambda\kappa} R_{\mu\nu} - g_{\mu\nu} R_{\lambda\kappa} +
		g_{\mu\kappa} R_{\lambda\nu}\right) + \frac{1}{6} R^{\alpha}_{\phantom{\alpha}\alpha} \left(g_{\lambda\nu} g_{\mu\kappa} - g_{\lambda\kappa} g_{\mu\nu}\right)
	\end{equation}
	is conformal. It is constructed by a particular combination of the Riemann tensor $R_{\lambda\mu\nu\kappa}$, Ricci tensor $R_{\mu\nu}$ and the Ricci scalar $R^{\alpha}_{\phantom{\alpha}\alpha}$. The~particular property of Weyl tensor is that it has the kinematic relation {$g_{\mu\kappa}C^{\lambda\mu\nu\kappa}=0$}. In~other words, the Weyl tensor is~traceless. 
	
	By imposing the principle of local conformal invariance as the requisite principle to restrict the choice of action for the gravitational field in curved spacetime, one requires the uniquely selected fourth-order gravitational action~\citep{1989ApJ...342..635M}
	\begin{equation}\label{eq:Weyl action}
		\begin{aligned}
			I_W =& -{\alpha}_g \int d^4x \sqrt{-g} C_{\lambda\mu\nu\kappa} C^{\lambda\mu\nu\kappa} \\
			=& -\alpha_g\int d^4x\sqrt{-g}\left[R_{\lambda\mu\nu\kappa}R^{\lambda\mu\nu\kappa}-2R_{\mu\nu}R^{\mu\nu}+(1/3){(R^\alpha_{\phantom{\alpha}\alpha})}^2\right] \\
			=& -2{\alpha}_g \int d^4x \sqrt{-g}\left[R_{\mu\nu}R^{\mu\nu}-(1/3){(R^\alpha_{\phantom{\alpha}\alpha})}^2\right] 
		\end{aligned}
	\end{equation}
	to remain invariant under any local metric transformation, where $\alpha_g$ is a dimensionless coupling constant. Variation of the action of Equation~(\ref{eq:Weyl action}) with respect to the metric yields
	\begin{equation}\label{eq:Weyl field tensor}
		\frac{1}{{(-g)}^{1/2}} \frac{\delta I_W}{\delta g_{\mu \nu}} = -2\alpha_gW^{\mu\nu}=-2\alpha_g\left[W_{(2)}^{\mu\nu}-\frac{1}{3}W_{(1)}^{\mu\nu}\right],
	\end{equation}
	where
	\begin{equation}\label{eq:sijiezhangliangfenshi}
		\begin{aligned}
			W^{\mu\nu}_{(1)} &= 2 g^{\mu\nu} (R^{\alpha}_{\phantom{\alpha}\alpha})^{;\beta}_{\phantom{;\beta};\beta} - 2 (R^{\alpha}_{\phantom{\alpha}\alpha})^{;\mu;\nu} - 2 R^{\alpha}_{\phantom{\alpha}\alpha} R^{\mu\nu} + \frac{1}{2} g^{\mu\nu} (R^{\alpha}_{\phantom{\alpha}\alpha})^2, \\
			W^{\mu\nu}_{(2)} &= \frac{1}{2} g^{\mu\nu} (R^{\alpha}_{\phantom{\alpha} \alpha})^{;\beta}_{\phantom{;\beta};\beta} + (R^{\mu\nu})^{;\beta}_{\phantom{;\beta};\beta} - (R^{\mu\beta})^{;\nu}_{\phantom{;\beta};\beta} - (R^{\nu\beta})^{;\mu}_{\phantom{;\beta} ;\beta} - 2 R^{\mu\beta} R^{\nu}_{\phantom{\beta}\beta} + \frac{1}{2} g^{\mu\nu} R^{\alpha\beta} R_{\alpha\beta}.
		\end{aligned} 
	\end{equation}
	
	{Conformal gravity requires the energy-momentum tensor $T_{\mu\nu}$ to be traceless, i.e., \mbox{$T^{\mu}_{\mu}=0$.} On~the other hand, elementary particle masses are not kinematic, but~rather that they are acquired dynamically by spontaneous breakdown. Hence, consider a massless, spin-$\frac{1}{2}$ matter field fermion $\psi(x)$, which is to obtain its mass through a massless, real spin-0 Higgs scalar boson field $S(x)$. The~required matter field action $I_M$ can be defined, as was done by ~\citep{mannheim_fitting_2012}}, as
	\begin{equation}\label{eq:IM}
		{ I_M = - \int d^4x \sqrt{-g} \left[ \frac{1}{2} S^{;\mu} S_{;\mu} - \frac{1}{12} S^2 R^\mu_{\phantom{\mu} \mu} + \lambda S^4 + i \bar{\psi} \gamma^\mu(x) \left( \partial_\mu + \Gamma_{\mu}(x) \right) \psi - h S \bar{\psi} \psi \right],}
	\end{equation}
	{where $h$ and $\lambda$ are dimensionless coupling constants, $\gamma^{\mu}(x)$ are the Dirac matrices, and $\Gamma_{\mu}(x)$ are the fermion spin connection.}
	Variation of the matter action $I_M$ with respect to the metric yields matter the source energy--momentum tensor $T^{\mu\nu}$
	\begin{equation}\label{eq:general energy momentum tensor}
		\begin{aligned}
			{ T^{\mu\nu} =} & {i\bar{\psi}\gamma^{\mu}(x)[\partial^\nu + \Gamma^{\nu}(x)]\psi + \frac{2}{3}S^{;\mu}S^{;\nu} - \frac{1}{6}g^{\mu\nu}S^{;\alpha}S_{;\alpha} - \frac{1}{3}S S^{;\mu;\nu} + \frac{1}{3}g^{\mu\nu}S S^{;\alpha}_{\phantom{;\alpha};\alpha}} \\
			& {- \frac{1}{6}S^2\left(R^{\mu\nu} - \frac{1}{2}g^{\mu\nu}R^{\alpha}_{\phantom{\alpha}\alpha}\right) - g^{\mu\nu}\left[\lambda S^4 + i\bar{\psi}\gamma^{\alpha}(x)[\partial_{\alpha}+\Gamma_{\alpha}(x)]\psi - hS\bar{\psi}\psi\right]. }  
		\end{aligned}
	\end{equation}

	The~variation of the total action $I_W+I_M$ with respect to the metric yields the equation of motion for CG
	\begin{equation}\label{eq:eom}
		4\alpha_gW^{\mu\nu}=T^{\mu\nu}.
	\end{equation}

	To date, however, the~exact solutions to Equation~(\ref{eq:eom}) can only be obtained for scenarios where $W^{\mu\nu}=T^{\mu\nu}=0$. The~difficulty stems from the fact that, in~CG, when $W^{\mu\nu}\ne 0$, we require an explicit dynamical model to describe how the gravitating system acquires its mass~\citep{1989ApJ...342..635M,mannheim_newtonian_1994}. Therefore, in~what follows, we will restrict our focus to the vacuum solutions where $T^{\mu\nu}=0$.
	
	To test Conformal Gravity (CG) using galaxy observations and gravitational lensing data, we must solve Equation~(\ref{eq:eom}) for both cosmological scenarios and a static, spherically symmetric~system. 
	
	In applying CG to cosmology, Weyl tensor vanishes in a Robertson--Walker metric~\citep{1992ApJ...391..429M}
	\begin{equation}
		ds^2 = c^2 dt^2 - R^2(t) \left[ \frac{dr^2}{1 - Kr^2} + r^2 d\theta^2 + r^2 \sin^2 \theta d\phi^2 \right].
	\end{equation}
	 {Thus} 
	$W^{\mu\nu}=0$, and~we can see from Equation~(\ref{eq:eom}) that $T^{\mu\nu}=0$. It turns out that conformal symmetry forbids the presence of any fundamental cosmological term, and~is thus a symmetry which is able to control the cosmological constant. {Even after the spontaneous breaking of conformal symmetry (which is required for particle mass generation), the~induced cosmological constant’s contribution to cosmology remains controlled~\citep{MANNHEIM2006340}. The~full content of the theory can be obtained by choosing a particular gauge in which the scalar field takes the constant value $S_0$. As~a result, the~energy--momentum tensor of Equation~(\ref{eq:general energy momentum tensor}) becomes~\citep{MANNHEIM2017125}}
	\begin{equation}\label{eq:cos energy momentum tensor}
		{ T^{\mu\nu} = i\bar{\psi}\gamma^{\mu}(x)\left[\partial^\nu + \Gamma^{\nu}(x)\right]\psi - \frac{1}{6} S_0^2\left(R^{\mu\nu} - \frac{1}{2}g^{\mu\nu}R^{\alpha}_{\phantom{\alpha}\alpha}\right) - g^{\mu\nu}\lambda S_0^4 = 0.}
	\end{equation}
	{{The averaging} of $i\bar{\psi}_{\mu}(x)[\partial + \Gamma_{\nu}(x)]\psi$ over all the fermionic modes propagating in a Robertson--Walker background reduces the fermionic contribution to $T^{\mu\nu}$, i.e., to that of a kinematic perfect~fluid}
	\begin{equation}\label{eq:perfect fluid}
		{T^{\mu\nu}_{\text{kin}} = \frac{1}{c} \left[(\rho_m + p_m) U^\mu U^\nu + p_m g^{\mu\nu}\right].}
	\end{equation}
	{Thus, the conformal cosmology equation of motion can be written as~\citep{MANNHEIM2006340}}
	\begin{equation}\label{eq:equation of motion 1}
		{ \frac{1}{6} S_0^2 \left(R^{\mu\nu} -  \frac{1}{2} g^{\mu\nu} R^\alpha_{\phantom{\alpha} \alpha} \right) 
			= \frac{1}{c} \left[(\rho_m + p_m) U^\mu U^\nu + p_m g^{\mu\nu}\right] - g^{\mu\nu} \lambda S_0^4.}
	\end{equation}
	 {When comparing} with the standard Einstein equation in $\Lambda$CDM model
	\begin{equation}\label{eq:LCDM Einstein equation}
		{-\frac{c^3}{8\pi G}\left(R^{\mu\nu}-\frac{1}{2}g^{\mu\nu}R^{\alpha}_{ \alpha}\right)=T^{\mu\nu}-g^{\mu\nu}\Lambda,}
	\end{equation}
	we only need to replace the gravitational constant $G$ by an effective, dynamically induced one $G_{\text{eff}} = - 3c^3/(4\pi S_0^2)$. We defined the conformal analogs of the standard $\Omega_M(t)$,  $\Omega_{\Lambda}(t)$ and $\Omega_K(t)$ via
	\begin{equation}
		\bar{\Omega}_M(t) = \frac{8\pi G_{\text{eff}} \rho_m(t)}{3c^2 H^2(t)}, \quad \bar{\Omega}_\Lambda(t) = \frac{8\pi G_{\text{eff}} \Lambda}{3c H^2(t)}, 
		\quad \bar{\Omega}_K(t) = -\frac{K c^2}{R^2(t) H^2(t)},
	\end{equation}
	where $H(t)=\dot{R(t)}/R(t)$ is the Hubble parameter, and $\Lambda = \lambda S_0^4$. As~usual, a Robertson--Walker geometry Equation~(\ref{eq:eom}) yields, at~redshift $z$, the~expression of the Hubble parameter 
	\begin{equation}
		{H(z) = H_0 \sqrt{\bar{\Omega}_{M0} {(1+z)}^3 + \bar{\Omega}_{K0} {(1+z)}^2 + \bar{\Omega}_{\Lambda 0}}},
	\end{equation}
	where {$\bar{\Omega}_{M0}=\bar{\Omega}_{M}(t=0)$}, etc. In~subsequent calculations, we adopted the values {$\bar{\Omega}_{K0}= 0.67$}, {$\bar{\Omega}_{\Lambda 0}=0.33$}, and~H$_0$ = 69.3 \,km s$^{-1}$  Mpc$^{-1}$, as~per reference~\citep{YANG201343}.
	
	For future reference, we defined the angular diameter distance as
	\begin{equation}\label{eq:angular diameter distance}
		{D_\text{A}(z_1,z_2)=\frac{1}{1+z_2}f_K\left[\chi(z_1,z_2)  \right],\quad f_K(\chi) = \frac{c}{H_0}\frac{1}{\sqrt{\bar{\Omega}_{K0}}}\sinh\left[\frac{H_0}{c}\sqrt{\bar{\Omega}_{K0}}\chi\right],}
	\end{equation}
	where
	\begin{equation}
		\chi(z_1,z_2) =  \int_{z_1}^{z_2} \frac{cdz'}{H(z')}.
	\end{equation}   
	 {The} proper distance is 
	\begin{equation}\label{eq:proper distance}
		{D^\text{P}(z_1,z_2)=\int_{z_1}^{z_2}\frac{cdz'}{(1+z')H(z')}.}
	\end{equation}

	For a static, spherically symmetric gravitational system, it turns out that the full kinematic content of CG is contained in the line element~\citep{PhysRevLett.53.315}
	\begin{equation}\label{eq:line element for a static spherical system}
		ds^2 = -B(r) dt^2 + \frac{dr^2}{B(r)} + r^2 (d\theta^2 + \sin^2 \theta d\phi^2).   
	\end{equation}
	 {Calculating} ${W^{\mu\nu}}$ {for this line element leads to}
	\begin{equation}\label{eq:combined solution}
		{  \frac{3}{B}(W^0_{\phantom{0}0} - W^r_{\phantom{r}r}) = B'''' + \frac{4 B'''}{r} = \frac{1}{r}(rB)'''' = \nabla^4 B.}
	\end{equation}
	{ {Defining} a source function}
	\begin{equation}\label{eq:f(r)}
		{ f(r) = \frac{3}{4\alpha_g B(r)} \left(T^0_{\phantom{0}0} - T^r_{\phantom{r}r} \right),}
	\end{equation}
	{the equations of motion of Equation~(\ref{eq:eom}) can be written as}
	\begin{equation}\label{eq:jianhuayundongfangcheng}
		{\nabla^4 B(r) = f(r).}
	\end{equation}
	 {The} exterior solution to Equation~(\ref{eq:jianhuayundongfangcheng}), with a radius of $r_0$, is {extremely important}
	for the applications of CG to strong lensing and galactic dynamics~\citep{1989ApJ...342..635M,MANNHEIM2006340}:
	\begin{equation}\label{eq:exterior solution}
		{ B(r > r_0) = -\frac{r}{2} \int_0^{r_0} dr' \, r'^2 f(r') - \frac{1}{6r} \int_0^{r_0} dr' \, r'^4 f(r') + w - \kappa r^2, }
	\end{equation}
	{where the $w - \kappa r^2$ term is the general solution to the homogeneous equation $\nabla^4 B(r) =0$. On~defining}
	\begin{equation}\label{eq:mkcanshu}
		{    \gamma = -\frac{1}{2} \int_0^{r_0} dr' \, r'^2 f(r') , \qquad 2\beta = \frac{1}{6} \int_0^{r_0} dr' \, r'^4 f(r'),}
	\end{equation}
	ignoring the $kr^2$ term, and setting $w=1$, the~metric of Equation~(\ref{eq:exterior solution}) can be written
	\begin{equation}\label{eq:final metric exterior to r0}
		B(r>r_0) = -g_{00}=\frac{1}{g_{rr}}=1-\frac{2\beta}{r}  + \gamma r.
	\end{equation}
	 {When} compared with the Schwarzschild solution in General Relativity (GR), the~$\beta$ term corresponds to the conventional Newtonian potential. The~$\gamma$ term, on~the other hand, represents an additional linear potential that is characteristic of Conformal Gravity (CG). The~presence of this $\gamma$ term enables CG to predict the rotation curves of spiral galaxies and the deflection angle of gravitational lensing without the need to invoke dark~matter.
	
	It is convenient to rewrite Equation~(\ref{eq:final metric exterior to r0}) in terms of the potential $V(r)$. The~rewritten form is as follows
	\begin{equation}\label{eq:metric in terms of V}
		B(r>r_0)=1+2V(r)/c^2, \ \ \text{with} \ \ V(r)=V_{\beta}+V_{\gamma} \ \ \text{and} \ \ V_{\beta}=-\frac{\beta c^2}{r}, \ \  V_{\gamma}=\frac{1}{2}\gamma c^2 r.
	\end{equation}
	In the region where $2\beta/r\gg\gamma r$, when $\beta=GM/c^2$, the~Schwarzschild solution $B(r>r_0)=1-\frac{2GM}{c^2r}$ can be recovered. Departures from this solution, specifically the linear potential $V_{\gamma}=\gamma c^2 r/2$, only occur at large distances. For~a typical star of the solar mass $M_{\odot}$, we write its potential as
	\begin{equation}\label{eq:potential of the sun}
		V^*(r)=-\frac{\beta^* c^2}{r} + \frac{\gamma^* c^2 r}{2},
	\end{equation}
	where $\beta^* = GM_\odot/c^2 = 1.48 \times 10^3$ m and $\gamma^*$ can be determined by observations. If~we denote $N^*=\frac{M}{M_{\odot}}$, $\beta=N^*\beta^*$, and {$\gamma = N^* \gamma^* + \gamma_0$}, then, for any point mass $M$, the~expression for its potential shown in Equation~(\ref{eq:metric in terms of V}) can be rewritten as
	\begin{equation}\label{eq:potential of any mass M}
		{
			V(r) = V_{\beta} + V_{\gamma} = -\frac{N^* \beta^* c^2}{r} + \frac{N^* \gamma^* c^2 r}{2}+\frac{\gamma_0 c^2 r}{2}.
		}
	\end{equation}
	 {Here}, \( \gamma_0 \) is a universal constant introduced through the fitting of spiral galaxy rotation curves. It does not depend on the mass distribution of any particular galaxy and is interpreted as the effective manifestation of the global geometry of the universe—especially the negative spatial curvature—within a local coordinate system. This corresponds to a universal linear gravitational potential term arising from the Hubble flow. Its physical origin reflects the global properties of the universe under the conformal metric, rather than the gravitational contribution from individual matter sources~\citep{Mannheim_1997}.

	We would like to point out that since $\gamma^*$ represents the linear potential associated with a unit of luminous mass, its value ought to be a universal constant. In~fact, through fitting the rotation curves of spiral galaxies, as~reported in reference~\citep{MANNHEIM2006340}, it has been found~that  
	\begin{equation}\label{eq:standard value of gamma*}
		{
			\gamma^\ast=5.42 \times 10^{-39} ~\text{m}^{-1}, \quad \gamma_0 = 3.06 \times 10^{-28} ~\text{m}^{-1}.
		}
	\end{equation}
	 {Furthermore}, as~demonstrated in the above analysis, \( \gamma_0 \) is regarded as a fixed constant across all galaxies. Throughout the remainder of this paper, we consistently adopt the value of \( \gamma_0 \) as specified earlier. Consequently, unless~otherwise noted, all references to the MK parameter in this work specifically refer to \( \gamma^* \).

	In the subsequent analysis, we aim to estimate the value of $\gamma^*$ using a statistic strong lensing approach.
	
	\section{Deflection Angle and Lensing~Equation} \label{sec:deflection angle}
	The deflection angle of the light path in a given metric of Equation~(\ref{eq:line element for a static spherical system}) around a point mass is derived by solving the null geodesic equation in the equatorial plane $\theta=\pi/2$ of the lensing object
	\begin{equation}\label{eq:geodesic}
		\frac{du}{d\phi} = \sqrt{\frac{1}{b^2} - B(u)u^2},
	\end{equation}
	{where \( u = \frac{1}{r} \), and~the integration constant \( b \) can be expressed in terms of the distance of closest approach \( r_0 \), which satisfies \( \left. \frac{dr}{d\psi} \right|_{r = r_0} = 0 \).}
	The function $B(r)$ is provided in the combined Equations~(\ref{eq:metric in terms of V}) and (\ref{eq:potential of any mass M}). The~deflection angle $\hat{\alpha}$ is derived by integrating $d\phi$ from the closest approach of the light path $r_0=1/u_0$ to the horizon $r_h=1/u_h$. The~closest point is defined by $\left. \frac{du}{d\phi}\right|_{u=u_0}=0$, and~the horizon is defined by $B(u_h)=0$. Thus, the deflection angle is given by
	\begin{equation}\label{eq:deflection angle of point lens}
		\hat{\alpha} = 2 \Delta \phi - \pi = 2 \int_{u_h}^{u_0} \frac{du}{\sqrt{\frac{1}{b^2} - u^2 B(u)}} - \pi.
	\end{equation}
	 {The} integral can be evaluated after a series expansion of the integrand to
	the second order in {$GMu_0^2/c^2$}. We, thus, obtain~\citep{PhysRevD.100.024019} 
	\begin{equation}\label{eq:kas et al}
		{
			\hat{\alpha} = \frac{4GM}{c^2r_0} + \frac{2GM\gamma}{c^2} + \left( \frac{15\pi}{4} - 4 \right) \frac{G^2M^2}{c^4r_0^2} + \left( \frac{15\pi}{4} - 4 \right) \frac{G^2M^2 \gamma}{c^4r_0}.
		}
	\end{equation}
	 {In} Conformal Gravity (CG), the~lensing equation has the same form as in General Relativity (GR). For~a point mass lens, we have
	\begin{equation}\label{eq:lens}
		\beta(\theta) = \theta - \frac{D_\text{LS}}{{D_\text{S}}} \hat{\alpha}(\theta),
	\end{equation}
	where $\beta(\theta)$, $\theta=r_0/D_\text{S}$, and $\hat{\alpha}(\theta)$ are the source position angle, image position angle, and
	deﬂection angle of Equation~(\ref{eq:kas et al}), respectively; $r_0$ is closest approach of light path from the lens; $D_\text{L}$, $D_\text{S}$, and $D_\text{LS}$ are the angular diameter distances from the observer to the lens, to~the source, and from the lens to the source, respectively. 
	{Specifically, by~setting \( \beta(\theta) = 0 \) in the lensing Equation~\eqref{eq:lens}, one can solve for the critical angular solution \( \theta \), which is known as the Einstein radius \( \theta_\mathrm{E} \). This angle corresponds to the formation of a ring-like image when the background source, the~lens, and~the observer are perfectly aligned, satisfying the critical condition for multiple imaging.}


	\section{\texorpdfstring{{Fitting the \boldmath$\gamma^*$ via Strong Gravitational~Lensing}}{Fitting the $\gamma^*$ via Strong Gravitational Lensing}}\label{sec:Fitting the gamma^*}
	
	{\citet{MANNHEIM2006340} determined the standard value of the linear potential parameter \( \gamma^* \) by fitting the rotation curves of spiral galaxies (see Equation~\eqref{eq:standard value of gamma*}). However, without~invoking dark matter, \citet{Ghosh_2023} applied this parameter to analyze the strong lensing phenomena observed in the galaxy clusters Abell 370 and Abell 2390. They found that CG, under~the standard value of \( \gamma^* \), fails to reproduce the observed lensing effects.
		This discrepancy suggests that the value of \( \gamma^* \) obtained from rotation curve fits may not be sufficient to account for strong gravitational lensing, thereby motivating the need to refit \( \gamma^* \) within lensing systems. As~can be seen from the deflection angle Formula \eqref{eq:kas et al} and the lensing Equation \eqref{eq:lens}, if~the observational data provide the redshifts of the background source \( z_S \) and the lensing object \( z_L \), the~Einstein radius \( \theta_{\mathrm{E}} \), and~the luminous mass enclosed within this radius \( M_\ast(\theta_{\mathrm{E}}) \), then the corresponding value of \( \gamma^* \) can be inferred.
		In this direction, related investigations have already been \mbox{conducted by \citet{Cutajar2014}.}}
	
	{
		After calculating the total linear potential parameter \( \gamma \) for each lensing system in their strong lensing galaxy cluster sample—where \( \gamma = N^* \gamma^* + \gamma_0 \)—they performed a fit to explore the correlation between \( \gamma \) and the gas mass \( M_\text{gas} \). The~results revealed a pronounced negative correlation between \( \gamma \) and \( M_\text{gas} \). More critically, the~magnitude of \( \gamma \) was found to be several orders of magnitude larger than the values obtained from rotation curve fits (for instance, the~minimum value of \( \gamma^* \) obtained from their fitting satisfied the relation \( \gamma^* = 0.33 \times (M/M_\odot)^{-1.164} \, \mathrm{m}^{-1} \)).
		However, it is worth noting that all deflection angle expressions adopted in that study omitted the crucial second-order term \( 2M\gamma \), as~compared to Equation~\eqref{eq:kas et al}. This omission could be a key factor contributing to the overestimated fit results.
		To further assess the impact of this missing term, we adopted the more complete deflection angle expression given by Equation~\eqref{eq:kas et al}, and~we refitted the relation between \( \gamma^* \) and the mass. The~use of \( \gamma^* \) here was intended to allow for a direct comparison with the values derived from rotation curve analyses.}
	
	{
		We performed the fitting using the strong gravitational lensing sample compiled by \citet{10.1093/mnras/stu106} (see Table~\ref{tab:strongGLtex}). This sample provides key observational quantities, including the effective radius \( \theta_e \), the~Einstein radius \( \theta_{\mathrm{E}} \), and~the total stellar mass \( M_* \), which are estimated under the assumption of a Salpeter initial mass function (IMF).
		Since the deflection angle \( \hat{\alpha} \) calculated in the point-mass model (see Equation~\eqref{eq:kas et al}) includes the Schwarzschild term \( 4M/r_0 \), where \( M \) denotes the stellar mass enclosed within radius \( r_0 \), we restricted our fitting of \( \gamma^* \) to the luminous mass enclosed within the Einstein radius, \( M_\ast( \theta_{\mathrm{E}} ) \). This approach is consistent with the treatment in \citet{Cutajar2014}, who also considered only the mass within the Einstein radius.
		It is important to emphasize that, unlike the Newtonian potential, the~linear potential term in Conformal Gravity exhibits a non-local character. Specifically, the~luminous mass located outside the Einstein radius \( \theta_{\mathrm{E}} \) does not produce mutually canceling gravitational effects at \( \theta_{\mathrm{E}} \); instead, it continues to contribute to light deflection. Therefore, by~fitting \( \gamma^* \) using only the mass within \( \theta_{\mathrm{E}} \), the~gravitational effects of the external mass are effectively absorbed into an elevated value of \( \gamma^* \), leading to an overestimation of this parameter.
		Nevertheless, such an overestimation does not alter the order of magnitude of \( \gamma^* \). Thus, the~fitted values obtained through this method remain valid for subsequent analyses.
	}
	
	\begin{table}[H]
		\small
		\caption{The original parameters and fitting parameters in strong gravitational lens systems. From Column 1 to Column 7: galaxy name, source redshift $z_S$, lens redshift $z_L$, effective angular radius $\theta_e$, Einstein radius $\theta_{\mathrm{E}}$, stellar mass $\log_{10}M_*(\theta_{\mathrm{E}})$, and the MK parameter $\gamma^*_\text{t}$. \label{tab:strongGLtex}}
		\begin{tabularx}{\textwidth}{cCCCCcc}
			\toprule
			\textbf{Name}       & \boldmath$z_L$ & \boldmath$z_S$ & \boldmath$\theta_e$    & \boldmath$\theta_{\mathrm{E}}$    & \boldmath$\log_{10}M_*(\theta_{\mathrm{E}})$    & \boldmath$\log_{10}\gamma^*_\text{t}$     \\
			&       &       & \textbf{[arcsec]}      & \textbf{[arcsec]}              &\textbf{ [\boldmath$\log_{10}(M_\odot)$]}        & \textbf{[\boldmath$\log_{10}$(m$^{-1}$)]}   \\
			\midrule
			SDSSJ0008$-$0004  & 0.440 & 1.192 & 1.71     & 1.16   & 11.11 & $-$30.85    \\
			SDSSJ0029$-$0055  & 0.227 & 0.931 & 2.16     & 0.96   & 10.93 & $-$31.00    \\
			SDSSJ0037$-$0942  & 0.195 & 0.632 & 1.80     & 1.53   & 11.36 & $-$31.73    \\
			SDSSJ0044$+$0113  & 0.120 & 0.197 & 1.92     & 0.80   & 10.80 & $-$30.53    \\
			SDSSJ0157$-$0056  & 0.513 & 0.924 & 1.84     & 0.79   & 11.00 & $-$30.66    \\
			SDSSJ0216$-$0813  & 0.332 & 0.523 & 2.40     & 1.16   & 11.38 & $-$31.27    \\
			SDSSJ0252$+$0039  & 0.280 & 0.982 & 1.39     & 1.04   & 11.01 & $-$30.95    \\
			SDSSJ0330$-$0020  & 0.351 & 1.071 & 0.91     & 1.10   & 11.24 & $-$31.48    \\
			SDSSJ0728$+$3835  & 0.206 & 0.688 & 1.78     & 1.25   & 11.20 & $-$31.51    \\
			SDSSJ0737$+$3216  & 0.322 & 0.581 & 1.80     & 1.00   & 11.36 & $-$31.74    \\
			SDSSJ0819$+$4534  & 0.194 & 0.446 & 1.98     & 0.85   & 10.79 & $-$30.51    \\
			SDSSJ0822$+$2652  & 0.241 & 0.594 & 1.82     & 1.17   & 11.19 & $-$31.28    \\
			SDSSJ0841$+$3824  & 0.116 & 0.657 & 4.21     & 1.41   & 10.91 & $-$30.73    \\
			SDSSJ0903$+$4116  & 0.430 & 1.065 & 1.78     & 1.29   & 11.35 & $-$31.35    \\
			SDSSJ0912$+$0029  & 0.164 & 0.324 & 4.01     & 1.63   & 11.22 & $-$30.90    \\
			SDSSJ0935$-$0003  & 0.347 & 0.467 & 2.15     & 0.87   & 11.27 & $-$31.01    \\
			SDSSJ0936$+$0913  & 0.190 & 0.588 & 2.11     & 1.09   & 11.08 & $-$31.39    \\
			SDSSJ0946$+$1006  & 0.222 & 0.609 & 2.35     & 1.38   & 11.04 & $-$30.68    \\
			SDSSJ0955$+$0101  & 0.111 & 0.316 & 1.47     & 0.91   & 10.48 & $-$29.83    \\
			SDSSJ0956$+$5100  & 0.241 & 0.470 & 2.19     & 1.33   & 11.24 & $-$31.05    \\
			SDSSJ0959$+$4416  & 0.237 & 0.531 & 1.98     & 0.96   & 11.07 & $-$31.14    \\
			SDSSJ0959$+$0410  & 0.126 & 0.535 & 1.29     & 0.99   & 10.67 & $-$30.40    \\
			SDSSJ1016$+$3859  & 0.168 & 0.439 & 1.46     & 1.09   & 11.01 & $-$31.03    \\
			SDSSJ1020$+$1122  & 0.282 & 0.553 & 1.59     & 1.20   & 11.31 & $-$31.36    \\
			SDSSJ1023$+$4230  & 0.191 & 0.696 & 1.77     & 1.41   & 11.12 & $-$31.06    \\
			SDSSJ1029$+$0420  & 0.104 & 0.615 & 1.56     & 1.01   & 10.75 & $-$31.15    \\
			SDSSJ1032$+$5322  & 0.133 & 0.329 & 0.81     & 1.03   & 10.80 & $-$30.48    \\
			SDSSJ1100$+$5329  & 0.317 & 0.858 & 2.20     & 1.52   & 11.35 & $-$31.31    \\
			SDSSJ1103$+$5322  & 0.158 & 0.735 & 2.85     & 1.02   & 10.73 & $-$30.45    \\
			SDSSJ1112$+$0826  & 0.273 & 0.629 & 1.32     & 1.49   & 11.39 & $-$31.44    \\
			SDSSJ1134$+$6027  & 0.153 & 0.474 & 2.02     & 1.10   & 10.89 & $-$30.75    \\
			SDSSJ1142$+$1001  & 0.222 & 0.504 & 1.24     & 0.98   & 11.13 & $-$31.46    \\
			SDSSJ1143$-$0144  & 0.106 & 0.402 & 2.66     & 1.68   & 11.14 & $-$31.19    \\
			SDSSJ1153$+$4612  & 0.180 & 0.875 & 1.16     & 1.05   & 10.95 & $-$31.16    \\
			SDSSJ1204$+$0358  & 0.164 & 0.631 & 1.09     & 1.31   & 11.12 & $-$31.35    \\
			SDSSJ1205$+$4910  & 0.215 & 0.481 & 1.79     & 1.22   & 11.22 & $-$31.28    \\
			SDSSJ1218$+$0830  & 0.135 & 0.717 & 2.70     & 1.45   & 11.03 & $-$31.00    \\
			SDSSJ1251$-$0208  & 0.224 & 0.784 & 2.61     & 0.84   & 10.72 & $-$30.45    \\
			SDSSJ1306$+$0600  & 0.173 & 0.472 & 1.25     & 1.32   & 11.11 & $-$31.03    \\
			SDSSJ1313$+$4615  & 0.185 & 0.514 & 1.59     & 1.37   & 11.18 & $-$31.20    \\
			SDSSJ1318$-$0313  & 0.240 & 1.300 & 2.51     & 1.58   & 11.18 & $-$31.09    \\
			SDSSJ1330$-$0148  & 0.081 & 0.711 & 0.96     & 0.86   & 10.28 & $-$29.72    \\
			SDSSJ1402$+$6321  & 0.205 & 0.481 & 2.29     & 1.35   & 11.22 & $-$31.17    \\
			SDSSJ1403$+$0006  & 0.189 & 0.473 & 1.14     & 0.83   & 10.98 & $-$32.13    \\
			SDSSJ1416$+$5136  & 0.299 & 0.811 & 0.98     & 1.37   & 11.34 & $-$31.46    \\
			SDSSJ1420$+$6019  & 0.063 & 0.535 & 2.25     & 1.04   & 10.52 & $-$30.56    \\
			SDSSJ1430$+$4105  & 0.285 & 0.575 & 2.55     & 1.52   & 11.34 & $-$31.16    \\
			\bottomrule
		\end{tabularx}
		
	\end{table}
	
	\begin{table}[H]\ContinuedFloat
		\small
		\caption{{\em Cont.}\label{tab:strongGLtex}}
		\begin{tabularx}{\textwidth}{cCCCCcc}
			\toprule
			\textbf{Name}       & \boldmath$z_L$ & \boldmath$z_S$ & \boldmath$\theta_e$    & \boldmath$\theta_{\mathrm{E}}$    & \boldmath$\log_{10}M_*(\theta_{\mathrm{E}})$    & \boldmath$\log_{10}\gamma^*_\text{t}$     \\
			&       &       & \textbf{[arcsec]}      & \textbf{[arcsec]}              &\textbf{ [\boldmath$\log_{10}(M_\odot)$]}        & \textbf{[\boldmath$\log_{10}$(m$^{-1}$)]}   \\
			\midrule
			SDSSJ1432$+$6317  & 0.123 & 0.664 & 3.04     & 1.26   & 11.03 & $-$31.74    \\
			SDSSJ1436$-$0000  & 0.285 & 0.805 & 1.63     & 1.12   & 11.17 & $-$31.26    \\
			SDSSJ1525$+$3327  & 0.358 & 0.717 & 2.42     & 1.31   & 11.38 & $-$31.35    \\
			SDSSJ1531$-$0105  & 0.160 & 0.744 & 1.97     & 1.71   & 11.32 & $-$31.68    \\
			SDSSJ1538$+$5817  & 0.143 & 0.531 & 1.00     & 1.00   & 10.92 & $-$31.46    \\
			SDSSJ1614$+$4522  & 0.178 & 0.811 & 2.58     & 0.84   & 10.63 & $-$30.35    \\
			SDSSJ1621$+$3931  & 0.245 & 0.602 & 1.51     & 1.29   & 11.32 & $-$31.55    \\
			SDSSJ1627$-$0053  & 0.208 & 0.524 & 1.98     & 1.23   & 11.11 & $-$31.00    \\
			SDSSJ1630$+$4520  & 0.248 & 0.793 & 1.65     & 1.78   & 11.47 & $-$31.66    \\
			SDSSJ1636$+$4707  & 0.228 & 0.675 & 1.68     & 1.08   & 11.15 & $-$31.53    \\
			SDSSJ1644$+$2625  & 0.137 & 0.610 & 1.55     & 1.27   & 11.01 & $-$31.20    \\
			SDSSJ1719$+$2939  & 0.181 & 0.578 & 1.46     & 1.28   & 11.06 & $-$30.97    \\
			SDSSJ2238$-$0754  & 0.137 & 0.713 & 1.82     & 1.27   & 10.96 & $-$31.03    \\
			SDSSJ2300$+$0022  & 0.228 & 0.463 & 1.52     & 1.24   & 11.19 & $-$31.05    \\
			SDSSJ2303$+$1422  & 0.155 & 0.517 & 2.94     & 1.62   & 11.14 & $-$30.98    \\
			SDSSJ2321$-$0939  & 0.082 & 0.532 & 4.11     & 1.60   & 10.89 & $-$30.72    \\
			SDSSJ2341$+$0000  & 0.186 & 0.807 & 2.36     & 1.44   & 11.18 & $-$31.31    \\
			SDSSJ2347$-$0005  & 0.417 & 0.714 & 1.14     & 1.11   & 11.46 & $-$31.61    \\
			SDSSJ0151$+$0049  & 0.517 & 1.364 & 0.67     & 0.75   & 11.00 & $-$30.97    \\
			SDSSJ0747$+$5055  & 0.438 & 0.898 & 1.09     & 0.64   & 10.93 & $-$30.88    \\
			SDSSJ0747$+$4448  & 0.437 & 0.897 & 0.92     & 0.72   & 10.93 & $-$30.71    \\
			SDSSJ0801$+$4727  & 0.483 & 1.518 & 0.50     & 0.89   & 10.97 & $-$30.75    \\
			SDSSJ0830$+$5116  & 0.530 & 1.332 & 0.97     & 0.89   & 11.03 & $-$30.81    \\
			SDSSJ0944$-$0147  & 0.539 & 1.179 & 0.48     & 0.92   & 11.10 & $-$30.88    \\
			SDSSJ1159$-$0007  & 0.579 & 1.346 & 0.96     & 0.81   & 11.04 & $-$30.87    \\
			SDSSJ1215$+$0047  & 0.642 & 1.297 & 0.65     & 0.74   & 11.33 & $-$32.01    \\
			SDSSJ1221$+$3806  & 0.535 & 1.284 & 0.47     & 0.74   & 11.05 & $-$31.07    \\
			SDSSJ1318$-$0104  & 0.659 & 1.396 & 0.69     & 0.84   & 11.18 & $-$31.10    \\
			SDSSJ1337$+$3620  & 0.564 & 1.182 & 2.03     & 0.68   & 10.92 & $-$30.66    \\
			SDSSJ1349$+$3612  & 0.440 & 0.893 & 1.89     & 0.71   & 10.81 & $-$30.38    \\
			SDSSJ1352$+$3216  & 0.463 & 1.034 & 0.58     & 0.86   & 11.10 & $-$31.02    \\
			SDSSJ1522$+$2910  & 0.555 & 1.311 & 0.89     & 0.74   & 10.95 & $-$30.74    \\
			SDSSJ1541$+$1812  & 0.560 & 1.113 & 0.76     & 0.93   & 11.09 & $-$30.77    \\
			SDSSJ1545$+$2748  & 0.522 & 1.289 & 2.59     & 0.42   & 10.53 & $-$30.29    \\
			SDSSJ1601$+$2138  & 0.543 & 1.446 & 0.44     & 0.91   & 11.19 & $-$31.28    \\
			SDSSJ1611$+$1705  & 0.477 & 1.211 & 1.00     & 0.74   & 10.75 & $-$30.28    \\
			SDSSJ1637$+$1439  & 0.391 & 0.874 & 1.04     & 0.75   & 10.73 & $-$30.21    \\
			SDSSJ2122$+$0409  & 0.626 & 1.452 & 0.90     & 0.63   & 10.83 & $-$30.52    \\
			SDSSJ2303$+$0037  & 0.458 & 0.936 & 1.35     & 0.39   & 10.69 & $-$31.41    \\
			HE0047$-$1756     & 0.408 & 1.670 & 0.49     & 0.80   & 10.93 & $-$30.93    \\
			Q0142$-$100       & 0.491 & 2.719 & 0.51     & 1.18   & 11.34 & $-$31.68    \\
			QJ0158$-$4325     & 0.317 & 1.294 & 0.66     & 0.58   & 10.65 & $-$30.72    \\
			HE0230$-$2130     & 0.522 & 2.162 & 0.14     & 0.87   & 10.75 & $-$30.25    \\
			SDSSJ0246$-$0825  & 0.723 & 1.686 & 0.18     & 0.53   & 10.93 & $-$31.05    \\
			MG0414$+$0534     & 0.958 & 2.639 & 0.78     & 1.11   & 11.44 & $-$31.49    \\
			HE0435$-$1223     & 0.454 & 1.689 & 0.76     & 1.22   & 11.12 & $-$30.91    \\
			B0712$+$472       & 0.406 & 1.339 & 0.36     & 0.72   & 10.92 & $-$31.02    \\
			MG0751$+$2716     & 0.350 & 3.200 & 0.31     & 0.42   & 9.92  & $-$29.02    \\
			HS0818$+$1227     & 0.390 & 3.115 & 0.62     & 1.37   & 11.03 & $-$30.71    \\
			B0850$+$054       & 0.588 & 3.930 & 0.16     & 0.34   & 10.44 & $-$30.85    \\
			SDSSJ0924$+$0219  & 0.393 & 1.523 & 0.30     & 0.88   & 10.91 & $-$30.74    \\
			LBQS1009$-$0252   & 0.871 & 2.739 & 0.19     & 0.77   & 10.78 & $-$30.24    \\
			J1004$+$1229      & 0.950 & 2.640 & 0.34     & 0.83   & 10.99 & $-$30.63    \\
			B1030$+$074       & 0.599 & 1.535 & 0.23     & 0.91   & 10.96 & $-$30.57    \\
			HE1104$-$1805     & 0.729 & 2.303 & 0.64     & 1.40   & 11.24 & $-$30.94    \\
			PG1115$+$080      & 0.311 & 1.736 & 0.46     & 1.14   & 10.93 & $-$30.65    \\
			RXJ1131$-$1231    & 0.295 & 0.658 & 1.13     & 1.83   & 11.09 & $-$30.46    \\
			SDSSJ1138$+$0314  & 0.445 & 2.442 & 0.19     & 0.57   & 10.77 & $-$31.15    \\
			SDSSJ1155$+$6346  & 0.176 & 2.888 & 0.43     & 0.76   & 10.62 & $-$30.83    \\
			SDSSJ1226$-$0006  & 0.517 & 1.126 & 0.45     & 0.57   & 10.90 & $-$30.95    \\
			\bottomrule
		\end{tabularx}
		
	\end{table}
	
	\begin{table}[H]\ContinuedFloat
		\small
		\caption{{\em Cont.}\label{tab:strongGLtex}}
		\begin{tabularx}{\textwidth}{cCCCCcc}
			\toprule
			\textbf{Name}       & \boldmath$z_L$ & \boldmath$z_S$ & \boldmath$\theta_e$    & \boldmath$\theta_{\mathrm{E}}$    & \boldmath$\log_{10}M_*(\theta_{\mathrm{E}})$    & \boldmath$\log_{10}\gamma^*_\text{t}$     \\
			&       &       & \textbf{[arcsec]}      & \textbf{[arcsec]}              &\textbf{ [\boldmath$\log_{10}(M_\odot)$]}        & \textbf{[\boldmath$\log_{10}$(m$^{-1}$)]}   \\
			\midrule
			LBQS1333$+$0113   & 0.440 & 1.571 & 0.31     & 0.85   & 10.91 & $-$30.72    \\
			Q1355$-$2257      & 0.702 & 1.373 & 1.24     & 0.62   & 10.91 & $-$30.63    \\
			HST14113$+$5211   & 0.465 & 2.811 & 0.47     & 0.84   & 10.67 & $-$30.16    \\
			HST14176$+$5226   & 0.809 & 3.400 & 0.70     & 1.41   & 11.04 & $-$30.50    \\
			B1422$+$231       & 0.337 & 3.620 & 0.32     & 0.78   & 10.65 & $-$30.34    \\
			MG1549$+$3047     & 0.111 & 1.170 & 0.82     & 1.15   & 10.69 & $-$30.47    \\
			B1608$+$656       & 0.630 & 1.394 & 0.64     & 0.81   & 11.36 & $-$32.00    \\
			PMNJ1632$-$0033   & 1.165 & 3.424 & 0.20     & 0.64   & 10.57 & $-$29.80    \\
			FBQ1633$+$3134    & 0.684 & 1.518 & 2.93     & 0.35   & 10.57 & $-$30.52    \\
			MG1654$+$1346     & 0.254 & 1.740 & 0.89     & 1.05   & 10.96 & $-$30.97    \\
			B1938$+$666       & 0.881 & 2.059 & 0.69     & 0.50   & 10.68 & $-$30.25    \\
			MG2016$+$112      & 1.004 & 3.273 & 0.22     & 1.78   & 11.35 & $-$30.96    \\
			WFI2033$-$4723    & 0.661 & 1.660 & 0.72     & 1.12   & 11.20 & $-$30.95    \\
			B2045$+$265       & 0.867 & 1.280 & 0.41     & 1.06   & 10.99 & $-$30.16    \\
			HE2149$-$2745     & 0.603 & 2.033 & 0.50     & 0.86   & 10.98 & $-$30.76    \\
			Q2237$+$030       & 0.039 & 1.695 & 3.86     & 0.90   & 10.03 & $-$29.31    \\
			COSMOS5921$+$0638 & 0.551 & 3.140 & 0.41     & 0.72   & 10.88 & $-$30.85    \\
			SDSSJ0743$+$2457  & 0.381 & 2.165 & 0.09     & 0.56   & 10.70 & $-$31.01    \\
			SDSSJ0806$+$2006  & 0.573 & 1.538 & 0.23     & 0.76   & 11.04 & $-$31.02    \\
			SDSSJ0819$+$5356  & 0.294 & 2.239 & 1.12     & 2.06   & 11.66 & $-$32.24    \\
			SDSSJ0946$+$1835  & 0.388 & 4.799 & 0.75     & 1.45   & 11.41 & $-$31.80    \\
			SDSSJ1055$+$4628  & 0.388 & 1.249 & 0.28     & 0.58   & 10.75 & $-$30.83    \\
			SDSSJ1620$+$1203  & 0.398 & 1.158 & 0.90     & 1.40   & 11.21 & $-$30.97    \\
			SL2SJ0213$-$0743  & 0.717 & 3.480 & 2.45     & 2.39   & 11.57 & $-$31.39    \\
			SL2SJ0214$-$0405  & 0.609 & 1.880 & 0.93     & 1.41   & 11.45 & $-$31.49    \\
			SL2SJ0217$-$0513  & 0.646 & 1.850 & 0.61     & 1.27   & 11.49 & $-$31.67    \\
			SL2SJ0219$-$0829  & 0.389 & 2.150 & 0.57     & 1.30   & 11.35 & $-$31.69    \\
			SL2SJ0225$-$0454  & 0.238 & 1.200 & 2.28     & 1.76   & 11.29 & $-$31.26    \\
			SL2SJ0226$-$0420  & 0.494 & 1.230 & 1.06     & 1.19   & 11.36 & $-$31.44    \\
			SL2SJ0232$-$0408  & 0.352 & 2.340 & 0.96     & 1.04   & 10.98 & $-$30.89    \\
			SL2SJ0849$-$0412  & 0.722 & 1.540 & 0.49     & 1.10   & 11.50 & $-$31.69    \\
			SL2SJ0849$-$0251  & 0.274 & 2.090 & 1.46     & 1.16   & 10.98 & $-$30.87    \\
			SL2SJ0901$-$0259  & 0.670 & 1.190 & 0.50     & 1.03   & 10.76 & $-$29.87    \\
			SL2SJ0904$-$0059  & 0.611 & 2.360 & 2.50     & 1.40   & 10.98 & $-$30.43    \\
			SL2SJ0959$+$0206  & 0.552 & 3.350 & 0.54     & 0.74   & 10.93 & $-$30.96    \\
			SL2SJ1359$+$5535  & 0.783 & 2.770 & 1.76     & 1.14   & 10.82 & $-$30.12    \\
			SL2SJ1405$+$5243  & 0.526 & 3.010 & 0.73     & 1.51   & 11.51 & $-$31.80    \\
			SL2SJ1406$+$5226  & 0.716 & 1.470 & 0.60     & 0.94   & 11.29 & $-$31.24    \\
			SL2SJ1411$+$5651  & 0.322 & 1.420 & 0.65     & 0.93   & 11.10 & $-$31.57    \\
			SL2SJ1420$+$5258  & 0.380 & 0.990 & 1.04     & 0.96   & 11.06 & $-$30.95    \\
			SL2SJ1420$+$5630  & 0.483 & 3.120 & 1.31     & 1.40   & 11.46 & $-$31.81    \\
			SL2SJ1427$+$5516  & 0.511 & 2.580 & 0.50     & 0.81   & 11.02 & $-$31.13    \\
			SL2SJ2203$+$0205  & 0.400 & 2.150 & 0.72     & 1.95   & 11.20 & $-$30.82    \\
			SL2SJ2213$-$0009  & 0.338 & 3.450 & 0.50     & 1.07   & 10.71 & $-$30.19    \\
			\bottomrule
		\end{tabularx}
	\end{table}
	
	{In the framework of Conformal Gravity, where dark matter is not introduced, we modeled the distribution of luminous matter in the lensing object using the Hernquist density profile. The~volume mass density was given by}
	\begin{equation}\label{eq:hernquist rho}
		\rho(r) = \frac{M_*}{2\pi} \frac{r_s}{r (r + r_s)^3},
	\end{equation}
	{where \( M_* \) denotes the total stellar mass of the system, and \( r_s \) is the scale radius, which is empirically related to the effective radius \( R_e \) of the lensing galaxy by \( r_s \approx 0.551 R_e \) \citep{1990ApJ...356..359H}.
		By adopting the transformation \( r = D_L \theta \), the~three-dimensional mass distribution can be recast in terms of angular coordinates. The~stellar mass enclosed within an angular radius \( \theta \) is then given by}
	\begin{equation}\label{eq:hernquist M}
		M(\theta) = \frac{\theta^2 M_*}{(\theta + \theta_s)^2},
	\end{equation}
	{where \( \theta_s = r_s / D_L \approx 0.551 \theta_e \) and~\( \theta_e \) denote the effective angular radius of the \mbox{lensing object.}}

\begin{figure}[H]
		\centering
		\includegraphics[width=8cm, angle=0]{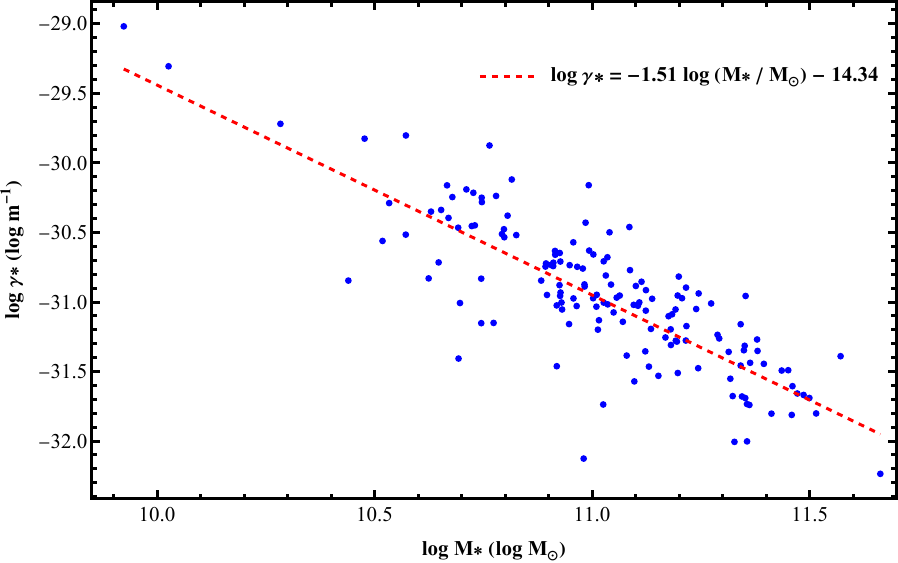}
		\caption{A plot showing the relation between the Mannheim–Kazanas linear potential parameter \( \gamma^* \) and the stellar mass \( M_*(\theta_\mathrm{E}) \) within the Einstein radius, which was derived from strong gravitational lensing observations. The~data points represent individual lensing systems, and~the dotted line denotes the empirical best-fit~result.}
		\label{fig:fig_2}
	\end{figure}

	{
		Using the observational data listed in Table~\ref{tab:strongGLtex}, we calculated the value of \( \gamma^* \) for each lensing system by substituting into the lensing Equation~\eqref{eq:lens} under the condition \( \beta = 0 \). Given that \( \gamma_0 \) is a universal constant determined by the cosmological background and remains the same across all systems, we fixed its value as \( \gamma_0 = 3.06 \times 10^{-28} \, \mathrm{m}^{-1} \).
		The computed values of \( \gamma^* \) were plotted in Figure~\ref{fig:fig_2}. Following the method adopted by \citet{Cutajar2014}, we performed a least-squares fit to examine the empirical relation between \( \gamma^* \) and the stellar mass \( M_*(\theta_\mathrm{E}) \) enclosed within the Einstein radius. The~resulting fitting formula was given by}
	\begin{equation}\label{eq：strong lens gamma*}
		\gamma^*_\text{t} = 4.57 \times 10^{-15} \left(\frac{M_*(\theta_\mathrm{E})}{M_\odot}\right)^{-1.51} \, \mathrm{m}^{-1}.
	\end{equation}
	{{To} 
		distinguish this result from the standard parameter value derived from spiral galaxy rotation curves, we denoted the fitting result from strong lensing as \( \gamma^*_\text{t} \). Compared to the findings of \citet{Cutajar2014}, our obtained \( \gamma^*_\text{t} \) was significantly lower, yet it still remained substantially higher than the value inferred from galactic rotation curve fitting.}

	{
		In the following section, we will perform a statistical analysis of the lensing probability distribution based separately on \( \gamma^* \) and \( \gamma^*_\text{t} \) in~order to further assess the explanatory power of different parameterizations with respect to the observational sample.
	}
	

	\section{Galaxy Stellar Mass~Function} \label{sec:GSMF}
	To obtain the lensing probability within the framework of Conformal Gravity (CG), the~galaxy stellar mass function (GSMF) is needed. It has been shown that, in~different redshift ranges, choosing either a Schechter function \citep{osti_7285770},
	\begin{equation}\label{eq:schechter1}
		\Phi(M)dM = \Phi_{\ast} {\left( \frac{M}{M_{\ast}} \right)}^{\alpha} \exp \left( - \frac{M}{M_{\ast}} \right) \frac{dM}{M_{\ast}} ,
	\end{equation}
	or a double Schechter function~\citep{Pozzetti2009zCOSMOS1},
	\begin{equation}\label{eq:schechter2}
		\Phi(M)dM = \left[ \Phi_{\ast1} {\left( \frac{M}{M_{\ast}} \right)}^{\alpha_1} + \Phi_{\ast2} {\left( \frac{M}{M_{\ast}} \right)}^{\alpha_2} \right] \exp \left( - \frac{M}{M_{\ast}} \right) \frac{dM}{M_{\ast}},
	\end{equation}
	can provide a good fit to the GSMF, where $\Phi(M)$ denotes the comoving number density of galaxies with mass between $M$ and $M+dM$,
	$\Phi_{\ast}$ is the normalization factor, and $M_{\ast}$ and $\alpha$ are characteristic stellar mass of galaxies and the low-mass slope, respectively.
	Relying on the photometric data in the near-IR bands of the COSMOS2015 catalog and correcting the Eddington bias,
	\citet{refId0} measured the GSMF in the redshift range of $0.2 \leq z \leq 5.5$. This was achieved using a Schechter function and double Schechter function, as~presented in Table~\ref{tab:GSMF}. For~galaxies with a redshift $z<0.2$, \citet{2009MNRAS.398.2177L} utilized a complete and uniform sample of 486,840 galaxies from the Sloan Digital Sky Survey to characterize the galaxy stellar mass function (GSMF) by fitting a Schechter function. They reported two sets of resulting parameters: one set was fitted in different stellar mass intervals, and~the other was fitted across the entire mass range. Although~the latter approach might overestimate the abundances of very massive galaxies, the~number density of such galaxies is extremely low. Consequently, it has little impact on the prediction of the lensing probability. For~convenience, in~this study, we adopted the GSMF fitted across the entire mass range to describe galaxies with $z<0.2$. The~specific parameter values are also presented in Table~\ref{tab:GSMF}.
	
	\begin{table}[H]
		\caption{The Schechter parameters across different redshift ranges. The parameter values for $z < 0.2$ are cited from \citet{2009MNRAS.398.2177L}. These values were obtained by fitting a Schechter function, and the reduced Hubble constant $h_{70} = 0.73$ was adopted. For the redshift range $0.2 < z \leq 5.5$, the parameter values were sourced from the reference~\citet{refId0}. Specifically, for $0.2 < z \leq 3.0$, the values were fitted using a double Schechter function, while, for $3.0 < z \leq 5.5$, a single Schechter function was used for fitting. In this case, the corresponding value of  $h_{70} = 1.0$.}
		\label{tab:GSMF}
		\begin{tabularx}{\textwidth}{LCccccccc}
			\toprule
			\textbf{Redshift} & \boldmath$\log M_{\ast}$          & \boldmath$\alpha_1$ & \boldmath$\Phi_{\ast1}$                  & \boldmath$\alpha_2$ & \boldmath$\Phi_{\ast2}$                   \\
			\midrule
			& \textbf{[\boldmath$h_{70}^{-2} M_\odot$]}  &            & \textbf{[\boldmath$10^{-3} h_{70}^{3} Mpc^{-3}$]} &            & \textbf{[\boldmath$10^{-3} h_{70}^{3} Mpc^{-3}$]}  \\
			\midrule 
			z $\leq$ 0.2 & 10.53 & $-$1.16 & 8.3   &  -    & -    \\
			0.2 < z $\leq$ 0.5 & 10.78 & $-$1.38 & 1.187 & $-$0.43 & 1.92 \\
			0.5 < z $\leq$ 0.8 & 10.77 & $-$1.36 & 1.070 & 0.03  & 1.68 \\
			0.8 < z $\leq$ 1.1 & 10.56 & $-$1.31 & 1.428 & 0.51  & 2.19 \\
			1.1 < z $\leq$ 1.5 & 10.62 & $-$1.28 & 1.069 & 0.29  & 1.21 \\
			1.5 < z $\leq$ 2.0 & 10.51 & $-$1.28 & 0.969 & 0.82  & 0.64 \\
			2.0 < z $\leq$ 2.5 & 10.60 & $-$1.57 & 0.295 & 0.07  & 0.45 \\
			2.5 < z $\leq$ 3.0 & 10.59 & $-$1.67 & 0.228 & $-$0.08 & 0.21 \\
			3.0 < z $\leq$ 3.5 & 10.83 & $-$1.76 & 0.090 & -     & -    \\
			3.5 < z $\leq$ 4.5 & 11.10 & $-$1.98 & 0.016 & -     &  -   \\
			4.5 < z $\leq$ 5.5 & 11.30 & $-$2.11 & 0.003 &   -   &  -   \\
			\bottomrule
		\end{tabularx}
	\end{table}

	\section{Lensing~Probability} \label{sec:lensing probability}
\begin{figure}[H]
		\centering
		\includegraphics[width=6cm, angle=-90]{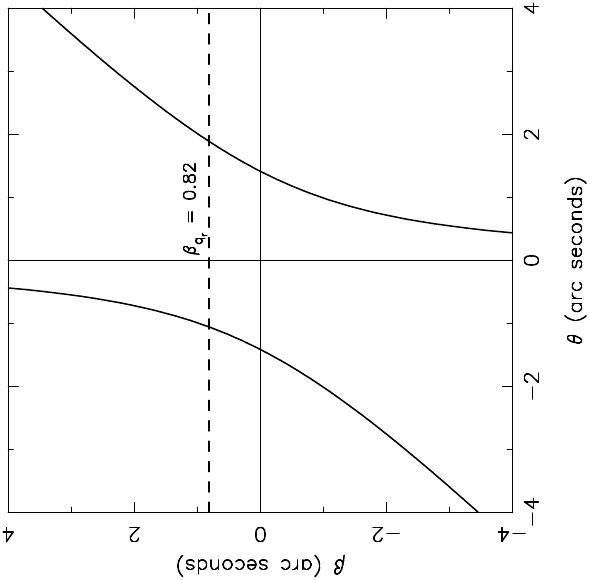}
		\caption{The lensing Equation~(\ref{eq:lens}) for a point luminous mass of $M=10^{11}M_\odot$. The~parameters were set as $\gamma^*=1.5 \times 10^{-31}$ m$^{-1}$, $z_S=1.57$, and $z_L=0.1$.  The~horizontal dashed line represents the allowed value of $\beta_{q_r}=0.82$, which is constrained by the largest flux density ratio $q_\text{r}=3.16$ of the sample that was used in ref.~\citep{Inada_2012}.} 
		\label{fig:lens eq beta-theta}
	\end{figure}

	As previously mentioned, within~the framework of Conformal Gravity (CG), we can only model the lens as a point mass. Despite the additional linear potential, namely the $\gamma$ term, the~lensing equation for a point mass in CG does not differ significantly from that in General Relativity (GR), as~can be seen from Equations~(\ref{eq:kas et al}) and (\ref{eq:lens}). In~Figure~\ref{fig:lens eq beta-theta}, we present a typical lensing equation for a point luminous mass of $M=10^{11}M_\odot$. The~source and the lens are located at redshifts $Z_S=1.57$ and $z_L=0.1$, respectively. As~depicted, analogous to the situation in GR, a~point-mass lensing system always has two images. We also found that, when $\theta\rightarrow 0$, $\beta(\theta)\rightarrow\infty$. Since the brightness of an image decreases as it gets closer to the lens and vice~versa,  the~flux density ratio $q_r$ between the two brighter and fainter images increases as $\beta$ increases. On~the other hand, any strong lensing sample has an allowed upper limit for $q_r$, which leads to a corresponding upper limit of $\beta_{q_r}$ for each lensing~system.
	
	\textls[-15]{In this study, we utilized the strong lensing sample presented in \mbox{reference \citet{Inada_2012}.}} In~this sample, the~magnitude difference between the two brighter and fainter images in the i-band was restricted to $\Delta i < 1.25$ mag, which implies $q_r = 10^{0.4 \times 1.25} \thickapprox  3.16$. The~magnifications of the two images are defined by
	\begin{equation}\label{eq:AmpliFact}
		\mu_{\pm} = {\left\lvert \frac{\theta}{\beta(\theta)}\frac{d\theta}{d\beta(\theta)} \right\rvert}_{\theta_\pm},
	\end{equation}
	where $\theta_\pm$ are the two solutions of the lensing equation $\beta(\theta)=0$, corresponding to the angular positions of the two images. The~flux density ratio $q_r$ is defined by $q_r = \mu_+/\mu_-$. Specifically, we have
	\begin{equation}\label{eq:AmpliFact2}
		\left( \frac{\theta(\beta)}{\beta} \frac{d\theta(\beta)}{d\beta} \right)_{\theta > 0}
		= q_r \left| \frac{\theta(\beta)}{\beta} \frac{d\theta(\beta)}{d\beta} \right|_{\theta_{0_-} < \theta < 0}.
	\end{equation}
	 {Here,} 
	$\theta_{0_-}$ represents the solution of the lensing equation $\beta(\theta) = 0$ on the negative half axis. Thus, $\beta_{q_r}$ can be obtained by solving the combined Equations~\eqref{eq:AmpliFact2} and ~\eqref{eq:lens}.

	{
		In addition, gravitational lensing can magnify the brightness of background sources, such as quasars or galaxies. As~a result, sources that would otherwise lie below the detection threshold (i.e., too faint to be observed) may be magnified above that threshold, thereby increasing their probability of being detected in survey samples. This leads to an observed number of lensing systems that exceeds the statistical expectation in the absence of lensing. On~the other hand, magnification also stretches the image of the background object, effectively enlarging the observed area on the sky and thus reducing the surface number density of sources per unit solid angle. The~net effect, known as magnification bias, is a combination of these two competing influences~\citep{1996astro.ph..6001N}.
		The expression for the magnification bias \( B(z_S, L) \) for a source at redshift \( z_S \) and intrinsic luminosity \( L \), under~the effect of gravitational lensing, is given by~\citep{1984ApJ...284....1T, Oguri_2002}:
		\begin{equation}\label{eq:AmplBias1}
			B(z_S, L) = \frac{2}{\eta_r^2 \, \Phi(z_S, L)} \int_0^{\eta_r} d\eta \, \eta \, \Phi\left(z_S, \frac{L}{\mu(\eta)}\right) \frac{1}{\mu(\eta)},
		\end{equation}
		where \( \Phi(z_S, L) \) is the luminosity function of the source population at redshift \( z_S \), \( \mu(\eta) \) is the lensing magnification factor, and~\( \eta_r \) characterizes the radial scale of the strong lensing region.
		A CLASS survey found that the number density of background sources follows a simple power--law relation \( \Phi(z_S, L) \propto L^{-\tilde{\gamma}} \). By~fitting the number–flux relation of flat-spectrum sources with flux densities in the range \( 30 \, \mathrm{mJy} < f < 200 \, \mathrm{mJy} \) at a frequency of \( 5 \, \mathrm{GHz} \), \citet{Rusin_2001} determined the best-fit slope to be \( \tilde{\gamma} = 2.07 \pm 0.11 \). In~general applications, this parameter is typically taken as \( \tilde{\gamma} = 2.1 \).
		With this simplification, Equation~\eqref{eq:AmplBias1} reduces to the following:
		\begin{equation}\label{eq:magnification bias}
			B(\beta) \approx \mu^{\tilde{ \gamma} - 1} \approx \mu^{1.1},
		\end{equation}
		where \( \mu = |\mu_-| + \mu_+ \) denotes the total magnification factor~\citep{1984ApJ...284....1T,Oguri_2002}.
	}

	The lensing cross section is a central concept in the theory of statistical gravitational lensing, and it is used to quantify the effective area covered by a lensing system under specific imaging conditions. More precisely, it describes the area on the source plane in which a source satisfies a given imaging criterion—such as the production of multiple images, an~image separation larger than a certain threshold, or~a flux ratio below a specified value—for a fixed lens–source geometry.
	The cross section on the source plane for systems producing image separations greater than \( \Delta \theta \) is defined as
	\( 
	\sigma_S(> \Delta \theta) = \frac{\pi \beta^2_{\mathrm{cr}}}{D_S^2} \, \Theta[\Delta\theta(M) - \Delta\theta],
	\)
	where \( \beta_{\mathrm{cr}} \) is the caustic radius on the source plane corresponding to multiple imaging, \( \Delta\theta(M) \) is the image separation produced by a lens with mass \( M \), and~\( \Theta(x) \) is the Heaviside function, which is defined as
	\[
	\Theta(x) =
	\begin{cases}
		1, & x \geq 0, \\
		0, & x < 0.
	\end{cases}
	\]
	 {For} practical purposes, the~lensing cross section is often transformed from the source plane to the lens plane. Using geometric relations, one obtains the following:
	\(
	\sigma(> \Delta \theta) = \sigma_S(> \Delta \theta) \, \frac{D_L^2}{D_S^2},
	\)
	which leads to the expression for the lensing cross section on the lens plane
	\begin{equation}\label{cross section}
		\sigma(> \Delta \theta) = \pi \beta^2_{\mathrm{cr}} D_L^2 \, \Theta[\Delta\theta(M) - \Delta\theta].
	\end{equation}
	 {This} expression is of fundamental importance in the statistical prediction of strong gravitational lensing~probabilities.

	A lensing cross section with an image separation larger than $\Delta\theta$ and a ﬂux
	density ratio less than $q_r$, when combined with the ampliﬁcation bias $B(\beta)$, is ~\citep{chen2006strong,Chen_2008}
	\begin{multline}\label{eq:sigma}
		\sigma_B(>\Delta \theta, \, <q_r) = 2\pi D_L^2 \times \\
		\hfill \begin{cases} 
			\int_0^{\beta_{q_r}} \beta \mu^{\tilde{\gamma} - 1}(\beta) \, d\beta, & \text{for } \Delta\theta \leq \Delta\theta_0, \\ 
			\left( \int_0^{\beta_{q_r}} - \int_0^{\beta_{\Delta\theta}} \right) \beta \mu^{\tilde{\gamma} - 1}(\beta) \, d\beta, & \text{for } \Delta\theta_0 < \Delta\theta \leq \Delta\theta_{q_r}, \\
			0, & \text{for } \Delta\theta > \Delta\theta_{q_r},
		\end{cases}
	\end{multline}  
	\textls[-25]{where $\beta_{\Delta\theta}$ is the source position at which a lens produces the image separation $\Delta\theta$;
	$\Delta\theta_0 = \Delta\theta(0)$} is the separation of the two images that are just on the Einstein ring; $\Delta\theta_{q_r}= \Delta\theta(\beta_{q_r})$ is the upper limit of the separation above which the ﬂux ratio of the two
	images will be greater than $q_r$; andc {\( \sigma_B \) is defined as the product of the lensing cross section \( \sigma \) and the magnification bias \( B(\beta) \).
	}
	
	At last, in~Conformal Gravity (CG), the~lensing probability for QSOs at a mean redshift $z_S$, which is lensed by foreground point-mass objects with~image separations greater than $\Delta\theta$ and flux density ratios below $q_r$, is given by
	\begin{equation}\label{eq:prob}
		P(>\Delta \theta, \, <q_r) = \int_0^{z_S} \frac{d D^p(z_L)}{dz_L} dz_L \int_{0}^\infty dM {(1+z_L)}^3 
		\phi(M,z_L) \sigma_B(>\Delta \theta, \, <q_r) ,
	\end{equation}
	where the proper distance $D^p(z_L)$ is provided in Equation~\eqref{eq:proper distance}, and the~comoving number density $\phi(M,z_L)$ can be selected from the galaxy stellar mass function (GSMF) fitted using a single Schechter function [Equation~\eqref{eq:schechter1}] or double Schechter function [Equation~\eqref{eq:schechter2}].
	
	We utilized the final statistical sample of gravitationally lensed quasars from the Sloan Digital Sky Survey (SDSS) Quasar Lens Search (SQLS)~\citep{Inada_2012}. This well-defined sample comprises 26 lensed quasars, with i-band magnitudes brighter than 19.1 and redshifts between 0.6 and 2.2, and they were selected from 50,826 spectroscopically confirmed quasars in SDSS Data Release 7 (DR7). The~sample is restricted to systems with image separations of 
	$1''< \Delta\theta < 20''$ and i-band magnitude differences of less than 1.25 mag between the two images. For~this lensed quasar sample, we adopted a mean source redshift of $z_S=1.57$  and constrained the flux density ratio to  $q_r = 10^{0.4 \times 1.25} \thickapprox  3.16$ (corresponding to i-band magnitude differences $< 1.25$ mag) in our lensing probability~calculations.

\begin{figure}[H]
		\centering
		\includegraphics[width=8cm, angle=-90]{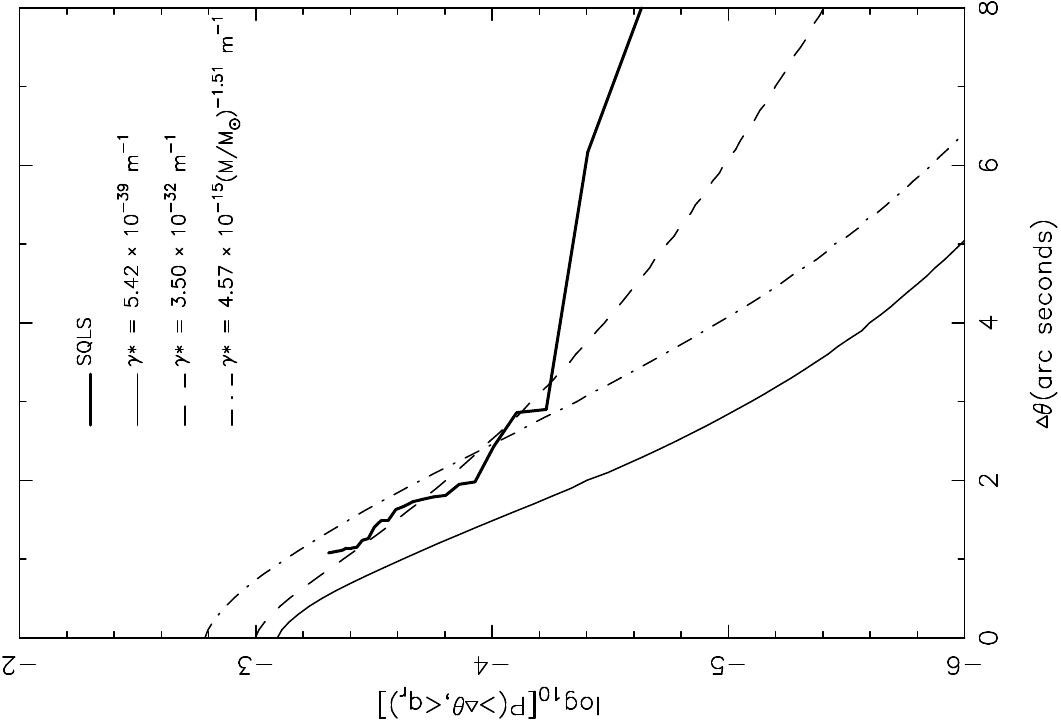}
		\caption{Lensing probability with image separations \( > \Delta \theta \) and flux density ratios \( < q_r = 3.16 \).  The~thick histogram shows the observed distribution from the~\citet{Inada_2012} sample. The~theoretical predictions are shown for the following: a standard rotation-curve-derived $\gamma^\ast=5.42 \times 10^{-39}$ m$^{-1}$ (thin solid line;~\citep{MANNHEIM2006340}); obtained from strong gravitational lensing fits $\gamma^*_\text{t} = 4.57 \times 10^{-15} \left(\frac{M_*(\theta_\mathrm{E})}{M_\odot}\right)^{-1.51}$ m$^{-1}$ (dash-dotted line); and a best-fit constant $\gamma^*_\text{c}=3.50\times 10^{-32}$ m$^{-1}$ (dashed line).}
		\label{fig:fig_3}
	\end{figure}

	The parameter $\gamma^*$ was constrained by fitting the lensing probability distribution predicted by Equation~(\ref{eq:prob}) to observations. Figure~\ref{fig:fig_3} presents the fitting results. The~observed probability (thick histogram) was derived from the~\citet{Inada_2012} sample through $P_{\text{obs}}(>\Delta \theta, \, <q_r) = N(>\Delta \theta, \, <q_r)/N_{\text{total}}$. The~thin solid line shows the lensing probability prediction using the rotation-curve-derived standard value $\gamma^\ast=5.42 \times 10^{-39}$ m$^{-1}$~\citep{MANNHEIM2006340}. This significantly underestimates the observed distribution (thick histogram), and it is consistent with single-cluster studies~\citep{Cutajar2014}.

	{
		The dot-dashed line represents the predicted probability distribution of strong gravitational lensing under Conformal Gravity when the MK parameter is set to the best-fit value \( \gamma^*_\text{t} \). It can be seen that, in~the region of small image separations \( \Delta \theta \lesssim 3'' \), the~theoretical prediction lies above the observed distribution.
		This behavior is consistent with our earlier analysis: since the point-mass lens model neglects the non-local contribution of stellar mass located outside the Einstein radius to the linear potential term, the~fitted value of \( \gamma^* \) is naturally elevated to compensate for the missing contribution. As~a result, the~predicted lensing probability is overestimated in the small-separation regime.
		In addition, a~more important factor is the negative correlation between \( \gamma^*_\text{t} \) and the lens mass: lower-mass lenses tend to yield larger fitted values of \( \gamma^*_\text{t} \). This enhances the contribution of the linear potential term in the deflection angle, increases the lensing cross section, and,~consequently, amplifies the lensing probability for low-mass systems in the small-separation regime.
		In contrast, for~image separations \( \Delta \theta \gtrsim 3'' \), the~theoretical prediction from Conformal Gravity falls significantly below the observed distribution. A~major reason for this discrepancy lies in the fact that the fitted \( \gamma^*_\text{t} \) decreases with increasing lens mass. For~massive galaxies or galaxy clusters, the~corresponding linear potential coefficient \( \gamma \) becomes smaller, thereby weakening the deflection angle and reducing the lensing~probability.
		
		For further comparison, we also plotted the predicted lensing probability distribution by assuming a constant \( \gamma^* \), which is shown as the dashed line in the figure. The~assumption of a constant \( \gamma^* \) is consistent with the fundamental premise of Conformal Gravity, wherein \( \gamma^* \) is taken to be the linear potential coefficient associated with a unit solar mass (for~a detailed discussion, see \citet{MANNHEIM2006340}).
		The best-fit constant MK parameter was found to~be
		\begin{equation}
			\gamma^*_\text{c} = 3.50 \times 10^{-32} \, \mathrm{m}^{-1}.
		\end{equation}
		 {As} 
		shown in Figure~\ref{fig:fig_3}, when the image separation is \( \Delta \theta \lesssim 2.5'' \), the~strong lensing probability predicted by Conformal Gravity with a constant MK parameter \( \gamma^*_\text{c} \) is lower than that obtained using the mass-dependent parameter \( \gamma^*_\text{t} \). In~contrast, for~\( \Delta \theta \gtrsim 2.5'' \), the~prediction based on \( \gamma^*_\text{c} \) exceeds that of \( \gamma^*_\text{t} \). This trend is consistent with our earlier analysis.
		However, when compared with the observational data, it is evident that Conformal Gravity underpredicts the strong lensing probability that occurs at large image separations, regardless of the form of \( \gamma^* \) adopted. Even when \( \gamma^* \) is tuned to its best-fit constant value from the data, the~current formulation of Conformal Gravity remains insufficient to fully account for the observed strong lensing probability~distribution.
		
		\citet{10.1093/mnras/stx1617} systematically investigated the strong lensing probability distributions corresponding to different dark matter halo density profiles within the framework of General Relativity. When using the same observational sample, SQLS~\citep{Inada_2012} (the~commonly used SIS+NFW composite model), as~was adopted in this study, showed good agreement with the observations in the small-separation regime, but~it significantly underestimated the occurrence rate of strong lensing events at large image separations. 
		This trend is consistent with our findings under the Conformal Gravity framework: whether adopting the mass-dependent MK parameter \( \gamma^*_\text{t} \) or the constant form \( \gamma^*_\text{c} \), the~theoretical predictions agree reasonably well with observations at small image separations, but they also, likewise, exhibit a noticeable underestimation in the large-separation~regime.
		
	}

	\section{Conclusions and~Discussions} \label{sec:conclusions}

	{
		\citet{Cutajar2014} derived an empirical relation between the total linear potential parameter \( \gamma \) (where \( \gamma = N^* \gamma^* + \gamma_0 \)) and the visible matter mass of the lens \( M_\ast(\theta_{\mathrm{E}}) \), and this was based on strong lensing observations of galaxy clusters. However, the~deflection angle formula employed in their analysis lacked the crucial second-order correction term \( 2m\gamma \), which likely led to an overestimation of the fitted results.
		To address this issue, we adopted a more accurate deflection angle expression \( \hat{\alpha} \), incorporated it into the lens equation, and~performed a new fit to the MK parameter \( \gamma^*_\text{t} \) using the strong lensing sample compiled by \citet{10.1093/mnras/stu106}. As~theoretically expected, the~resulting value of \( \gamma^*_\text{t} \) was significantly lower than that reported by \citet{Cutajar2014}, yet it remained several orders of magnitude higher than the standard \( \gamma^* \) value obtained from galaxy dynamics in the non-relativistic limit.
		This result is consistent with previous findings based on galaxy clusters, such as Abell 370 and Abell 2390~\citep{Ghosh_2023}.
		Moreover, the~parameter exhibited a clear mass dependence, and it decreased with increasing lens mass.

		Subsequently, we performed, considering different values of the MK parameter \( \gamma^* \), a statistical analysis of the lensing probability distribution using the expression given in Equation~\eqref{eq:prob}. The~results were then compared with the complete strong lensing observational sample provided by \citet{Inada_2012}.
		Our analysis shows that, when \( \gamma^* \) is fixed at the standard value obtained from fitting the rotation curves of spiral galaxies, the~theoretical prediction falls significantly below the observational data across the entire range of image separations. As~a result, it fails to effectively reproduce the strong lensing events observed in reality.
		This finding suggests that, although~Conformal Gravity performs well in explaining the dynamical behavior of spiral galaxies—thereby successfully reproducing the observations without invoking dark matter—it encounters difficulties in remaining consistent with strong lensing observations when assuming a constant linear potential parameter.
		This also poses certain challenges to the universality of Conformal Gravity across different physical scales and phenomena, suggesting that its current formulation may be insufficient to fully accommodate the diverse observational constraints ranging from internal galactic dynamics to gravitational~lensing.
		
		When the MK parameter is taken as the mass-dependent value \( \gamma^*_\text{t} \) that is obtained from strong gravitational lensing fits, the~prediction slightly exceeds the observational results in the image separation range \( \Delta \theta \lesssim 3'' \). This overestimation arises from two key factors: first, \( \gamma^*_\text{t} \) decreases with increasing lens mass; second, the~point-mass lens model neglects the contribution of matter outside the Einstein radius to the linear gravitational potential at \( \theta_\text{E} \).
		However, as~the image separation increases, the~theoretical prediction rapidly declines and falls significantly below the observed values, thereby failing to account for the strong lensing events at large separations.
		In addition, we obtained a best-fit constant parameter \( \gamma^*_\text{c} \) from the observational data, 
		{which is consistent with the requirement that 
			\( \gamma^*\) be a constant, as determined from the galaxy rotation curve fits by \citet{MANNHEIM2006340}.} The fitted value of \( \gamma^*_\text{c} \) exceeded the standard value derived from spiral galaxy rotation curves by approximately seven orders of magnitude. Although~this constant improved the prediction in certain parameter ranges, it still failed to reproduce the observed statistics at large image separations.
		This result indicates that even adopting a constant \( \gamma^* \) is insufficient to resolve the discrepancy currently faced by Conformal Gravity in the context of strong lensing statistics.
		This conclusion is consistent with the predictions made under General Relativity when assuming a dark matter halo distribution modeled by the SIS+NFW combination~\citep{10.1093/mnras/stx1617}.

		In summary, our results indicate that Conformal Gravity faces certain difficulties in providing a self-consistent explanation for multi-scale observational phenomena, ranging from galactic dynamics to strong gravitational~lensing. 
		
		Moreover, the~quantum gravity theory proposed by \citet{universe10080333,Chen_2022} also incorporates a linear potential term similar to that in Conformal Gravity. A~systematic investigation of this theory at galactic scales and in the context of gravitational lensing may provide further insights into the viability of explaining observational phenomena without invoking dark matter through linear gravitational potentials. Such studies would also allow for a comparative assessment of the performance of linear potentials across different gravitational frameworks, potentially offering new perspectives for understanding the fundamental nature of gravity.
	}
	
	\vspace{6pt} 
	\authorcontributions{Conceptualization, L.-X.Y.; Methodology, L.-X.Y.; Software, L.-X.Y.; Validation, L.-X.Y.; Formal analysis, L.-X.Y.; Investigation, L.-X.Y.; Writing---original draft, D.-M.C.; Writing---review \& editing, D.-M.C. All authors have read and agreed to the published version of \mbox{the manuscript.}}

	\funding{This work was supported by the National Natural Science Foundation of China (Grant No. 12203066).}

	\dataavailability{The data used in this study are entirely obtained from previously published sources, as~cited in the references. No new data were generated or analyzed in this~study.}
	
	\acknowledgments{We would like to sincerely thank the anonymous referees for their careful reading of our manuscript and for providing many insightful comments and constructive suggestions. Their efforts have greatly contributed to improving the clarity, accuracy, and~overall quality of this~work.} 
	
	\conflictsofinterest{The authors declare no conflicts of~interest.}

	\begin{adjustwidth}{-\extralength}{0cm}
		
	\reftitle{References}


\begin{thebibliography}{999}

\bibitem[Schneider et~al.(1999)Schneider, Ehlers, and
Falco]{schneider1999gravitational}
Schneider, P.; Ehlers, J.; Falco, E.
\newblock {\em Gravitational Lenses}; Astronomy and Astrophysics Library;
Springer:   {Berlin/Heidelberg, Germany,} 
1999.

\bibitem[Weinberg(1972)]{Weinberg:1972kfs}
Weinberg, S.
\newblock {\em {Gravitation and Cosmology}: {Principles and Applications of the
General Theory of Relativity}}; John Wiley and Sons: \mbox{New York}, NY, USA, 1972.

\bibitem[Mannheim(2006)]{MANNHEIM2006340}
Mannheim, P.D.
\newblock Alternatives to dark matter and dark energy.
\newblock {\em Prog. Part. Nucl. Phys.} {\bf 2006}, {\em 56},~340--445. [\href{http://doi.org/10.1016/j.ppnp.2005.08.001}{CrossRef}]

\bibitem[{Mannheim}(1992)]{1992ApJ...391..429M}
{Mannheim}, P.D.
\newblock {Conformal Gravity and the Flatness Problem}.
\newblock {\em Astrophys. J.} {\bf 1992}, {\em 391},~429. [\href{http://dx.doi.org/10.1086/171358}{CrossRef}]

\bibitem[{Mannheim}(2000)]{Mannheim2000FoPh}
{Mannheim}, P.D.
\newblock {Attractive and repulsive gravity.}
\newblock {\em Found. Phys.} {\bf 2000}, {\em 30},~709--746.
 [\href{http://dx.doi.org/10.1023/A:1003737011054}{CrossRef}]

\bibitem[Mannheim(2001)]{Mannheim_2001}
Mannheim, P.D.
\newblock Cosmic Acceleration as the Solution to the Cosmological Constant
Problem.
\newblock {\em  Astrophys. J.} {\bf 2001}, {\em 561},~1. [\href{http://dx.doi.org/10.1086/323206}{CrossRef}]

\bibitem[Yang et~al.(2013)Yang, Chen, Zhao, Li, and Liu]{YANG201343}
Yang, R.; Chen, B.; Zhao, H.; Li, J.; Liu, Y.
\newblock Test of conformal gravity with astrophysical observations.
\newblock {\em Phys. Lett. B} {\bf 2013}, {\em 727},~43--47. [\href{http://dx.doi.org/10.1016/j.physletb.2013.10.035}{CrossRef}]

\bibitem[Mannheim and Kazanas(1989)]{1989ApJ...342..635M}
Mannheim, P.D.; Kazanas, D.
\newblock Exact vacuum solution to conformal Weyl gravity and galactic rotation
curves.
\newblock {\em Astrophys. J.} {\bf 1989}, {\em 342},~534--544. [\href{http://dx.doi.org/10.1086/167623}{CrossRef}]

\bibitem[Mannheim and O’Brien(2012)]{mannheim_fitting_2012}
Mannheim, P.D.; O’Brien, J.G.
\newblock Fitting galactic rotation curves with conformal gravity and a global
quadratic potential.
\newblock {\em Phys. Rev. D} {\bf 2012}, {\em 85},~124020. [\href{http://dx.doi.org/10.1103/PhysRevD.85.124020}{CrossRef}]

\bibitem[Mannheim and O'Brien(2013)]{Mannheim_2013}
Mannheim, P.D.; O'Brien, J.G.
\newblock Galactic rotation curves in conformal gravity.
\newblock {\em J. Physics Conf. Ser.} {\bf 2013}, {\em
437},~012002. [\href{http://dx.doi.org/10.1088/1742-6596/437/1/012002}{CrossRef}]

\bibitem[{O'Brien} and {Moss}(2015)]{2015JPhCS.615a2002O}
{O'Brien}, J.G.; {Moss}, R.J.
\newblock {Rotation curve for the Milky Way galaxy in conformal gravity}.
In \emph{Journal of Physics: Conference Series,  {Proceedings of the 9th Biennial Conference on Classical and Quantum Relativistic Dynamics of Particles and Fields (IARD 2014), Storrs, CT, USA, 9–13 June 2014}
}; IOP Publishing: Bristol, UK, 2015; Volume 615, p. 012002. [\href{http://dx.doi.org/10.1088/1742-6596/615/1/012002}{CrossRef}]

\bibitem[Edery and Paranjape(1998)]{PhysRevD.58.024011}
Edery, A.; Paranjape, M.B.
\newblock Classical tests for Weyl gravity: Deflection of light and time delay.
\newblock {\em Phys. Rev. D} {\bf 1998}, {\em 58},~024011. [\href{http://dx.doi.org/10.1103/PhysRevD.58.024011}{CrossRef}]

\bibitem[Sultana and Kazanas(2010)]{PhysRevD.81.127502}
Sultana, J.; Kazanas, D.
\newblock Bending of light in conformal Weyl gravity.
\newblock {\em Phys. Rev. D} {\bf 2010}, {\em 81},~127502. [\href{http://dx.doi.org/10.1103/PhysRevD.81.127502}{CrossRef}]

\bibitem[Sultana(2013)]{sultana2013deflection}
Sultana, J.
\newblock Deflection of light to second order in conformal Weyl gravity.
\newblock {\em J. Cosmol. Astropart. Phys.} {\bf 2013}, {\em
2013},~048. [\href{http://dx.doi.org/10.1088/1475-7516/2013/04/048}{CrossRef}]

\bibitem[Cattani et~al.(2013)Cattani, Scalia, Laserra, Bochicchio, and
Nandi]{PhysRevD.87.047503}
Cattani, C.; Scalia, M.; Laserra, E.; Bochicchio, I.; Nandi, K.K.
\newblock Correct light deflection in Weyl conformal gravity.
\newblock {\em Phys. Rev. D} {\bf 2013}, {\em 87},~047503. [\href{http://dx.doi.org/10.1103/PhysRevD.87.047503}{CrossRef}]

\bibitem[Lim and Wang(2017)]{PhysRevD.95.024004}
Lim, Y.-K.; Wang, Q.-h.
\newblock Exact gravitational lensing in conformal gravity and
Schwarzschild--de Sitter spacetime.
\newblock {\em Phys. Rev. D} {\bf 2017}, {\em 95},~024004. [\href{http://dx.doi.org/10.1103/PhysRevD.95.024004}{CrossRef}]

\bibitem[Ka{\c{s}}{\i}kc{\i} and Deliduman(2019)]{PhysRevD.100.024019}
Ka{\c{s}}{\i}kc{\i}, O.; Deliduman, C.
\newblock Gravitational lensing in Weyl gravity.
\newblock {\em Phys. Rev. D} {\bf 2019}, {\em 100},~024019. [\href{http://dx.doi.org/10.1103/PhysRevD.100.024019}{CrossRef}]

\bibitem[Mannheim and Kazanas(1994)]{mannheim_newtonian_1994}
Mannheim, P.D.; Kazanas, D.
\newblock Newtonian limit of conformal gravity and the lack of necessity of the
second order {Poisson} equation.
\newblock {\em Gen. Relativ. Gravit.} {\bf 1994}, {\em
26},~337--361. [\href{http://dx.doi.org/10.1007/BF02105226}{CrossRef}]

\bibitem[Mannheim(2017)]{MANNHEIM2017125}
Mannheim, P.D.
\newblock Mass generation, the cosmological constant problem, conformal
symmetry, and the Higgs boson.
\newblock {\em Prog. Part. Nucl. Phys.} {\bf 2017}, {\em
94},~125--183. [\href{http://dx.doi.org/10.1016/j.ppnp.2017.02.001}{CrossRef}]

\bibitem[Riegert(1984)]{PhysRevLett.53.315}
Riegert, R.J.
\newblock Birkhoff's Theorem in Conformal Gravity.
\newblock {\em Phys. Rev. Lett.} {\bf 1984}, {\em 53},~315--318. [\href{http://dx.doi.org/10.1103/PhysRevLett.53.315}{CrossRef}]

\bibitem[Mannheim(1997)]{Mannheim_1997}
Mannheim, P.D.
\newblock Are galactic rotation curves really flat?
\newblock {\em  Astrophys. J.} {\bf 1997}, {\em 479},~659--664. [\href{http://dx.doi.org/10.1086/303933}{CrossRef}]

\bibitem[Ghosh et~al.(2023)Ghosh, Bhattacharya, Sherpa, and Bhadra]{Ghosh_2023}
Ghosh, S.; Bhattacharya, M.; Sherpa, Y.; Bhadra, A.
\newblock Test of conformal theory of gravity as an alternative paradigm to
dark matter hypothesis from gravitational lensing studies.
\newblock {\em J. Cosmol. Astropart. Phys.} {\bf 2023}, {\em 2023},~ {008.} 
\newblock [\href{http://dx.doi.org/10.1088/1475-7516/2023/07/008}{CrossRef}]

\bibitem[Cutajar and Zarb~Adami(2014)]{Cutajar2014}
Cutajar, D.; Zarb~Adami, K.
\newblock Strong lensing as a test for conformal Weyl gravity.
\newblock {\em Mon. Not. R. Astron. Soc.} {\bf 2014}, {\em 441},~1291--1296. [\href{http://dx.doi.org/10.1093/mnras/stu617}{CrossRef}]

\bibitem[Oguri et~al.(2014)Oguri, Rusu, and Falco]{10.1093/mnras/stu106}
Oguri, M.; Rusu, C.E.; Falco, E.E.
\newblock The stellar and dark matter distributions in elliptical galaxies from
the ensemble of strong gravitational lenses.
\newblock {\em Mon. Not. R. Astron. Soc.} {\bf 2014},
{\em 439},~2494--2504. [\href{http://dx.doi.org/10.1093/mnras/stu106}{CrossRef}]

\bibitem[{Hernquist}(1990)]{1990ApJ...356..359H}
\textls[-15]{{Hernquist}, L.
\newblock {An Analytical Model for Spherical Galaxies and Bulges}.
\newblock {\em Astrophys. J.} {\bf 1990}, {\em 356},~359. [\href{http://dx.doi.org/10.1086/168845}{CrossRef}]}

\bibitem[Schechter(1976)]{osti_7285770}
Schechter, P.
\newblock An analytic expression for the luminosity function for galaxies.
\newblock {\em Astrophys. J.} {\bf 1976}, {\em 203},~297--306. [\href{http://dx.doi.org/10.1086/154079}{CrossRef}]

\bibitem[{Pozzetti, L.} et~al.(2010){Pozzetti, L.}, {Bolzonella, M.}, {Zucca,
E.}, {Zamorani, G.}, {Lilly, S.}, {Renzini, A.}, {Moresco, M.}, {Mignoli,
M.}, {Cassata, P.}, {Tasca, L.}, {Lamareille, F.}, {Maier, C.}, {Meneux, B.},
{Halliday, C.}, {Oesch, P.}, {Vergani, D.}, {Caputi, K.}, {Kovač, K.},
{Cimatti, A.}, {Cucciati, O.}, {Iovino, A.}, {Peng, Y.}, {Carollo, M.},
{Contini, T.}, {Kneib, J.-P.}, {Le Févre, O.}, {Mainieri, V.}, {Scodeggio,
M.}, {Bardelli, S.}, {Bongiorno, A.}, {Coppa, G.}, {de la Torre, S.}, {de
Ravel, L.}, {Franzetti, P.}, {Garilli, B.}, {Kampczyk, P.}, {Knobel, C.}, {Le
Borgne, J.-F.}, {Le Brun, V.}, {Pellò, R.}, {Perez Montero, E.},
{Ricciardelli, E.}, {Silverman, J. D.}, {Tanaka, M.}, {Tresse, L.}, {Abbas,
U.}, {Bottini, D.}, {Cappi, A.}, {Guzzo, L.}, {Koekemoer, A. M.}, {Leauthaud,
A.}, {Maccagni, D.}, {Marinoni, C.}, {McCracken, H. J.}, {Memeo, P.},
{Porciani, C.}, {Scaramella, R.}, {Scarlata, C.}, and {Scoville,
N.}]{Pozzetti2009zCOSMOS1}
{Pozzetti, L.}; {Bolzonella, M.}; {Zucca, E.}; {Zamorani, G}.; {Lilly,
S.}; {Renzini, A.}; {Moresco, M.}; {Mignoli, M.}; {Cassata, P.}; {Tasca,
L.};  et~al.
\newblock zCOSMOS – 10k-bright spectroscopic sample---The bimodality in the
galaxy stellar mass function: Exploring its evolution with redshift.
\newblock {\em Astron. Astrophys.} {\bf 2010}, {\em 523},~A13. [\href{http://dx.doi.org/10.1051/0004-6361/200913020}{CrossRef}]

\bibitem[Davidzon et~al.(2017)Davidzon, Ilbert, Laigle, Coupon, McCracken,
Delvecchio, Masters, Capak, Hsieh, Le~Fèvre, Tresse, Bethermin, Chang,
Faisst, Le~Floc’h, Steinhardt, Toft, Aussel, Dubois, Hasinger, Salvato,
Sanders, Scoville, and Silverman]{refId0}
Davidzon, I.; Ilbert, O.; Laigle, C.; Coupon, J.; McCracken, H.J.; Delvecchio,
I.; Masters, D.; Capak, P.; Hsieh, B.C.; Le~Fèvre, O.;  et~al.
\newblock The COSMOS2015 galaxy stellar mass function: Thirteen billion years
of stellar mass assembly in ten snapshots.
\newblock {\em Astron. Astrophys.} {\bf 2017}, {\em 605},~A70. [\href{http://dx.doi.org/10.1051/0004-6361/201730419}{CrossRef}]

\bibitem[{Li} and {White}(2009)]{2009MNRAS.398.2177L}
{Li}, C.; {White}, S.D.M.
\newblock {The distribution of stellar mass in the low-redshift Universe}.
\newblock {\em Mon. Not. R. Astron. Soc.} {\bf 2009}, {\em 398},~2177--2187. [\href{http://dx.doi.org/10.1111/j.1365-2966.2009.15268.x}{CrossRef}]

\bibitem[Inada et~al.(2012)Inada, Oguri, Shin, Kayo, Strauss, Morokuma, Rusu,
Fukugita, Kochanek, Richards, Schneider, York, Bahcall, Frieman, Hall, and
White]{Inada_2012}
Inada, N.; Oguri, M.; Shin, M.S.; Kayo, I.; Strauss, M.A.; Morokuma, T.; Rusu,
C.E.; Fukugita, M.; Kochanek, C.S.; Richards, G.T.;  et~al.
\newblock The Sloan Digital Sky Survey Quasar Lens Search. V. Final Catalog
from the Seventh Data Release.
\newblock {\em Astron. J.} {\bf 2012}, {\em 143},~119. [\href{http://dx.doi.org/10.1088/0004-6256/143/5/119}{CrossRef}]

\bibitem[{Narayan} and {Bartelmann}(1996)]{1996astro.ph..6001N}
{Narayan}, R.; {Bartelmann}, M.
\newblock Lectures on Gravitational Lensing.
\newblock {\em arXiv} {\bf 1996}, arXiv:astro-ph/9606001.

\bibitem[Turner et~al.(1984)Turner, Ostriker, and Gott]{1984ApJ...284....1T}
Turner, E.L.; Ostriker, J.P.; Gott, J.R.I.
\newblock The statistics of gravitational lenses: The distributions of image
angular separations and lens redshifts.
\newblock {\em Astrophys. J.} {\bf 1984}, {\em 284},~1--22. [\href{http://dx.doi.org/10.1086/162379}{CrossRef}]

\bibitem[Oguri et~al.(2002)Oguri, Taruya, Suto, and Turner]{Oguri_2002}
Oguri, M.; Taruya, A.; Suto, Y.; Turner, E.L.
\newblock Strong Gravitational Lensing Time Delay Statistics and the Density
Profile of Dark Halos.
\newblock {\em Astrophys. J.} {\bf 2002}, {\em 568},~488--499. [\href{http://dx.doi.org/10.1086/339064}{CrossRef}]

\bibitem[Rusin and Tegmark(2001)]{Rusin_2001}
Rusin, D.; Tegmark, M.
\newblock Why Is the Fraction of Four-Image Radio Lens Systems So High?
\newblock {\em Astrophys. J.} {\bf 2001}, {\em 553},~709. [\href{http://dx.doi.org/10.1086/320955}{CrossRef}]

\bibitem[Chen and Zhao(2006)]{chen2006strong}
Chen, D.M.; Zhao, H.
\newblock Strong lensing probability for testing TeVeS theory.
\newblock {\em  Astrophys. J.} {\bf 2006}, {\em 650},~L9. [\href{http://dx.doi.org/10.1086/508612}{CrossRef}]

\bibitem[Chen(2008)]{Chen_2008}
Chen, D.M.
\newblock Strong lensing probability in TeVeS (tensor--vector--scalar) theory.
\newblock {\em J. Cosmol. Astropart. Phys.} {\bf 2008}, {\em 2008},~006. [\href{http://dx.doi.org/10.1088/1475-7516/2008/01/006}{CrossRef}]

\bibitem[Wang et~al.(2017)Wang, Chen, and Li]{10.1093/mnras/stx1617}
Wang, L.; Chen, D.M.; Li, R.
\newblock Baryon effects on the dark matter haloes constrained from strong
gravitational lensing.
\newblock {\em Mon. Not. R. Astron. Soc.} {\bf 2017},
{\em 471},~523--531. [\href{http://dx.doi.org/10.1093/mnras/stx1617}{CrossRef}]

\bibitem[Chen and Wang(2024)]{universe10080333}
Chen, D.M.; Wang, L.
\newblock Quantum Effects on Cosmic Scales as an Alternative to Dark Matter and
Dark Energy.
\newblock {\em Universe} {\bf 2024}, {\em 10},~333. [\href{http://dx.doi.org/10.3390/universe10080333}{CrossRef}]

\bibitem[Chen(2022)]{Chen_2022}
Chen, D.M.
\newblock Torsion Fields Generated by the Quantum Effects of Macro-bodies.
\newblock {\em Res. Astron. Astrophys.} {\bf 2022}, {\em
22},~125019. [\href{http://dx.doi.org/10.1088/1674-4527/ac9aef}{CrossRef}]
			
		\end{thebibliography}

		\PublishersNote{}
	\end{adjustwidth}
\end{document}